\documentclass[usegraphicx, usenatbib]{mn2e}
\usepackage{times}
\usepackage[usenames, dvipsnames]{color}
\usepackage{graphicx}
\usepackage{pdflscape}
\usepackage{amsmath}
\usepackage{rotating}
\usepackage{natbib}
\usepackage{amssymb}
\usepackage{tablefootnote}
\newlength{\colwidth}
\renewcommand{\footnotesize}{\scriptsize}

\title[Optical Polarization Catalogue of Blazars]
  {The RINGO2 and DIPOL Optical Polarisation Catalogue of Blazars}
\author[H. Jermak et al.]
  {H.~Jermak,$^1$
I. A. ~Steele,$^1$
E. ~Lindfors,$^2$ 
T. ~Hovatta,$^{3,4}$ 
K.~Nilsson,$^5$ 
G. P. Lamb,$^1$
C. ~Mundell,$^6$
\newauthor  
U. ~Barres de Almeida,$^7$ 
A. ~Berdyugin, $^2$
V. Kadenius, $^2$
R. Reinthal,$^2$
L. Takalo, $^2$
             \\
  $^1$Astrophysics Research Institute, Liverpool John Moores University, Brownlow Hill, Liverpool, UK, L3 5RF.\\
$^2$Tuorla Observatory, Department of Physics and Astronomy, University of Turku, V\"ais\"al\"antie 20, 21500 Piikki\"o, Finland.\\
$^3$Aalto University Mets\"ahovi Radio Observatory, Mets\"ahovintie 114, 02540 Kylm\"al\"a, Finland.\\
$^4$Aalto University Department of Radio Science and Engineering,P.O. BOX 13000, FI-00076 AALTO, Finland. \\
$^5$Finnish Center for Astrophysics with ESO, University of Turku, V\"ais\"al\"antie 20, 21500 Piikki\"o, Finland.\\
$^6$Department of Physics, Bath University, Bath, UK, BA2 7AY.\\
$^7$Centro Brasileiro de Pesquisas Fisicas, Rua Dr. Xavier Sigaud 150, Urca, Rio de Janeiro, RJ 22290-160, Brazil. \\}
\date{Released 2016 Xxxxx XX}
\pagerange{\pageref{firstpage}--\pageref{LastPage}} \pubyear{2014}

\def\LaTeX{L\kern-.36em\raise.3ex\hbox{a}\kern-.15em
    T\kern-.1667em\lower.7ex\hbox{E}\kern-.125emX}

\begin{document}

\label{firstpage}

\maketitle

\begin{abstract}
We present $\sim$2000 polarimetric and $\sim$3000 photometric observations of 15 $\gamma$-ray bright blazars over a period of 936 days (11/10/2008 - 26/10/2012) using data from the Tuorla blazar monitoring program (KVA DIPOL) and Liverpool Telescope (LT) RINGO2 polarimeters (supplemented with data from SkyCamZ (LT) and Fermi-LAT $\gamma$-ray data). In 11 out of 15 sources we identify a total of 19 electric vector position angle (EVPA) rotations and 95 flaring episodes. We group the sources into subclasses based on their broadband spectral characteristics and compare their observed optical and $\gamma$-ray properties. We find that (1) the optical magnitude and $\gamma$-ray flux are positively correlated, (2) EVPA rotations can occur in any blazar subclass, 4 sources show rotations that go in one direction and immediately rotate back, (3) we see no difference in the $\gamma$-ray flaring rates in the sample; flares can occur during and outside of rotations with no preference for this behaviour, (4) the average degree of polarisation (DoP), optical magnitude and $\gamma$-ray flux are lower during an EVPA rotation compared with during non-rotation and the distribution of the DoP during EVPA rotations is not drawn from the same parent sample as the distribution outside rotations, (5) the number of observed flaring events and optical polarisation rotations are correlated, however we find no strong evidence for a temporal association between individual flares and rotations and (6) the maximum observed DoP increases from $\sim$10\% to $\sim$30\% to $\sim$40\% for subclasses with synchrotron peaks at high, intermediate and low frequencies respectively.
\end{abstract}


\section{Introduction}
The centre of most, if not all galaxies, contains at least one supermassive black hole \citep{kormendy1995,magorrian1998}. If the matter in the vicinity of the compact object is close enough to become accreted onto the compact object then it is classified as an active galactic nucleus (AGN).
The viewing angle of an AGN often determines its observational classification. Those AGN viewed within a small opening angle of the jet axis are classified as blazars \citep{urrypadovani1995}. Blazars are defined by rapid flux variability with large amplitudes, high apparent luminosities, greater brightness temperatures than typical AGN, high polarisation and superluminal motion of ejected components in the jet. The apparent superluminary properties are caused by the relativistic beaming of the jet emission towards the observer \citep{blandfordrees1978}.

Questions of the formation, collimation and acceleration of blazar jets from the regions close to the supermassive black hole are still unsolved, however progress can be made by exploring the signatures of the magnetic field in polarised light. Using the linear Stokes Parameters to calculate the angle and degree of polarisation we can explore how the optical synchrotron emission evolves during a $\gamma$-ray flare and whether rotations in the electric vector position angle (EVPA) correspond with low- or high- states in the optical and $\gamma$-ray emission. Changes in the EVPA and the degree of polarisation can afford information about the structure and order of the underlying magnetic field \citep{kikuchi1988}.

Blazars are the most energetic of the AGN classes and have characteristic \textquoteleft double-humped\textquoteright \:spectral energy distributions (SEDs) that span the entire electromagnetic spectrum. The first hump, peaking in the Infrared-optical is attributed to synchrotron emission, whereas the higher energy peak (X-ray to $\gamma$-ray) is thought to be produced by Inverse- or synchrotron self-Compton scattering of jet or external photons. Blazars can be sub-divided according to the location of the synchrotron emission peak in their spectral energy distribution (SED). Flat spectrum radio quasars (those sources originally identified to have optical emission line equivalent widths $\geq$5\AA) are included with BL Lac objects that have SED synchrotron peaks at low-frequencies (low-synchrotron-peaked LSP) ($\nu$ \textless10$^{14}$ Hz (IR), $\lambda\gtrsim$30000 \AA). The intermediate synchrotron peaked sources (ISPs) and high synchrotron peaked sources (HSPs) are all BL Lac objects and have synchrotron peaks between 10$^{14}$ $\leq \nu \leq$ 10$^{15}$ (optical/IR, $\lambda\sim$3000-30000 \AA) and $\geq$10$^{15}$Hz (UV, $\lambda\lesssim$3000 \AA) respectively \citep{abdo2010b}. 

Due to the variability of blazars, the subclass of FSRQ/BL~Lac is a loose one. The optical emission spectrum of a source has been shown to vary between the two classes depending on the activity state of the source (e.g. \cite{ghisellini2011}). For this reason in this comparison we mainly focus on studying the blazar subclasses according to the literature location in the spectral energy distribution of their synchrotron peak. With the peak being directly related to the energy distribution of the electrons within the jet this classification allows the analysis of sources which may have different high energy emission processes.

The high energy component of the SED can be explained by synchrotron self-Compton (SSC) emission \citep{marscher1985,maraschi1992,bloom1996,tavecchio1998}, a combination of leptonic SSC and hadronic synchrotron emission (e.g. \cite{mannheim1992,mucke2001,bottcher2013}), synchrotron emission of leptons produced in a hadronic cascade via proton-photon collisions above the pion production energy (e.g. \cite{mannheim1992, aharonian2000, mucke2001, mucke2003, bottcher2013}), or external Compton (EC) of soft seed photons from a variety of sources such as: accretion disc radiation \citep{dermer1992, dermer1993}); optical and ultraviolet emission from the circumnuclear material \citep{sikora1994, blandford1995, ghisellini1996, dermer1997}; infra-red emission from the dusty torus \citep{blazejowski2000}; or synchrotron emission from other jet regions \citep{georganopoulos2003, ghisellini2008}. 

%Hadronic models also suggest interactions between protons at relativistic energies result in pion production. These pions decay to $\gamma$-rays as $\pi^0 \rightarrow$ 2 $\gamma$  \citep{mannheim1993}. 

%In hadronic models the high energy emission consists of synchrotron emission from protons, protons collide to produce pions, $\gamma$-ray photons (from pion decay  $\pi^0 \rightarrow$ 2 $\gamma$ \citep{mannheim1993}), synchrotron and Compton emission from secondary decay products of charged pions and the output from pair cascades initiated by these high-energy emissions instrinsically absorbed by $\gamma\gamma$ pair production \citep{mannheim1992,aharonian2000,mucke2001,mucke2003,bottcher2013}. As in leptonic models, in these scenarios the lower frequency hump is caused by synchrotron radiation.

Some proposed causes of polarisation angle rotations within blazar jets are emission features or shocks travelling along helical paths or magnetic field lines (e.g.\cite{marscher2008,marscher2010,zhang2014}, a turbulent magnetic field resulting in the random walk of the polarisation vector \citep{jones1985,marscher2014}. Visual rather than physical effects such as a bent jet and the trajectory of the polarisation angle on the sky can also cause polarisation angle rotations \citep{abdo2010}. \cite{young2010} graphically presents the idea shown by \cite{bjornsson1982} that other kinematic effects causing variations of the viewing angle in the co-moving frame could explain EVPA rotations up to 180$^\circ$. \cite{nalewajko2010} continue with this idea and suggest that a symmetric emitting region on a bent jet could produce a gradual EVPA rotation.

%Testing the correlation of optical to $\gamma$-ray flares can be used to constrain the various models. \cite{bottcher2010} demonstrated that the lag/lead time of optical to $\gamma$-ray emission can be extremely sensitive to the external photon density, although the lag/lead time range was found to be relatively short (a few hours). \cite{ghisellini1996} predict a $\gamma$-ray flare to lag optical emission if seed photons are mirrored by clouds within the broad line region (BLR), whereas \cite{tavani2015} have recently shown that if no correlation is evident between $\gamma$-ray and optical flares then a mirror outside of the BLR can explain such events.

In this paper we present the results of a polarimetric and photometric campaign on a sample of fifteen blazars and present the correlations between these data and Fermi $\gamma$-ray data. We also explore the relationship between $\gamma$-ray flares and polarisation angle rotations. In \textsection2 we present the photometric and polarimetric data analysis procedures and the definitions of $\gamma$-ray flares and EVPA rotations used in this analysis. In \textsection3 we detail the sample and the historical behaviour of the blazars. In \textsection 4 we discuss the results of the correlation analysis and the differences between blazar subclasses according to the location of the synchrotron peak in the spectral energy distributions. In \textsection5 we present and discuss our findings from the analysis. In the Appendix we present the light curves of each of the sources in this sample. The data presented in the publication will be made available via the Vizier source \footnote{http://vizier.u-strasbg.fr/viz-bin/VizieR}.

\section{Observations}
The polarisation monitoring program with RINGO2 is a continuation of a program that was running at the KVA-60 telescope in 2009-2011. The KVA-60 telescope is used for optical support observations of the MAGIC telescopes and has a relatively small mirror diameter of 60 cm. The source sample originally consisted of 8 $\gamma$-ray bright blazars that had an optical magnitude of R\textless16 and were known to show strong (\textgreater5\%) polarisation: 3C~66A, S5~0716+714, OJ~287, ON~231, 3C~279, PKS~1510-089, PG~1553+113 and BL~Lac. For these 8 sources we have the longest polarisation light curves. When the monitoring with RINGO2 started in 2010, the sample was gradually expanded with AGN the MAGIC Cherenkov telescope \citep{lorenz2005} has been following: which were Mrk~421, Mrk~501 (long-term monitoring programs), 1ES~1011+496, PKS~1222+216, Mrk~180 (short term multi-wavelength campaigns and target of opportunity observations). See Table \ref{RINGO2_sample} for the full sample of fifteen sources.

%\begin{table*}
%\begin{tabular}{l c c c c c}
%\hline
%Target & RA (J2000)& Dec (J2000) & R Mag. & z &Data unavailable\\
%\hline
%1ES 1426+428 & 14:28:32.6 & +42:40:21 & 14.4 & 0.129&Fermi$^1$, KVA (pol)\\
%1ES 1218+304 & 12:21:21.9 & +30:10:36.8 & 15.7 & 0.164&KVA (pol.)\\
%1ES 1011+496 & 10:15:04.19& +49:26:01.0& 15.4 &  0.212&KVA (pol)\\
%PKS 1222+216 & 12:24:54.40 &+21:22:46.0& 15.6 &   0.432&\\
%PG 1553+113 & 15:55:43.00& +11:11:24.0&     13.8 &  \textless0.78&\\
%BL Lac & 22:02:43.28&		+42:16:40.0&     14.5 &  0.069&\\
%S5 0716+714 & 07:21:53.45&+71:20:36.4&      13.7 &    0.31 &\\
%ON 231 & 12:21:31.69&		+28:13:58.5&    15    & 0.102&\\
%Mrk 501 &16:53:52.20&		+39:45:37.0&     13.2 &  0.034&KVA (pol)\\
%OJ 287 & 08:54:48.90&		+20:06:30.9&    15.2  & 0.306&\\
%Mrk 180 & 11:36:26.40&		+70:09:27.0&     14.6 &  0.045&Fermi$^1$, KVA (pol)\\
%3C 279 & 12:56:11.17&		-05:47:21.5&    16    & 0.536&SkyCamZ$^2$\\
%Mrk 421 & 11:04:27.30&		+38:12:32.0&     13   &  0.031&\\
%PKS~1510-089-089& 15:12:50.53& -09:05:59.8&   16   &   0.360&\\
%3C 66A &02:22:39.61&		+43:02:07.8&     14.6 &  0.444&\\
%\hline
%\item $^1$Unavailable $^2$Poor data quality
%\end{tabular}
%\caption{The full RINGO2 catalogue with RA, Dec. and redshift information. References for the redshift values can be found in the Appendix. The %missing data is detailed in the table along with an explanation for its absence.}\label{RINGO2_sample}
%\end{table*}

% \usepackage{graphicx}
\begin{table*}%[h]
\begin{minipage}{\textwidth}    
%\begin{threeparttable}
\centering
\resizebox{\textwidth}{!}{%
\begin{tabular}{|l|l|l|l|l|l|l|l|}
\hline
{\bf Name}       & {\bf z }     &  {\bf Type}   & {\bf R Mag. range}&{\bf Pol. range (\%)}&{\bf Fermi range}& {\bf  Observation Period (MJD)} &  {\bf Absent data}       \\ \hline
 {\bf3C~66A      }& 0.444       & ISP    & 15.1-13.2  &    1.0-27.7 & 4.0x10$^{-8}$-1.1x10$^{-6}$  & 55413.17 - 56226.03    & ...                \\ %\hline
{\bf S5~0716+714 }& 0.31  & ISP    & 14.4-12.2  &    0.3-23.7  &6.4x10$^{-8}$-1.3x10$^{-6}$& 55651.86 - 56035.00    & ...                 \\ %\hline
{\bf OJ~287      }& 0.306       & LSP & 15.4-13.5  &  4.5-38.7&2.8x10$^{-8}$-1.4x10$^{-6}$ &55641.91 - 56223.22    & ...                \\ %\hline
 {\bf 1ES~1011+496    }& 0.212     & HSP    & 15.6-14.7 & 0.8-6.8 &4.2x10$^{-8}$- 9.7x10$^{-8}$& 56006.93 - 56094.92    & KVA (pol) \\ %\hline
 {\bf Mrk~421}    & 0.031     &   HSP     &    13.0-12.0     & 0.2-8.8 & 1.0x10$^{-7}$-1.3x10$^{-6}$  & 55705.90 - 56096.89   & RINGO2              \\ %\hline
 {\bf Mrk~180}& 0.045      &  HSP     &15.5-15.1 & 2.5-5.1& ...& 56006.89 - 56216.24                      & Fermi, KVA (pol)   \\ %\hline
{\bf 1ES~1218+304    }& 0.164         & HSP    & 16.0-15.5         & 0.6-4.3&1.6x10$^{-8}$- 2.6x10$^{-8}$ & 56065.88 - 56136.90    & KVA (pol) \\ %\hline
 {\bf ON~231     } & 0.102     & ISP    & 15.6-14.1  &  0.6-23.3  &2.3x10$^{-8}$-8.6x10$^{-8}$& 55573.26 - 56032.97    & ...                 \\ %\hline
 {\bf PKS~1222+216    }& 0.432     & LSP   & 15.8-14.7   &  0.5-9.7  &8.6x10$^{-8}$-1.3x10$^{-6}$ &55901.24 - 55935.16    &...                 \\ %\hline
 {\bf 3C~279      }& 0.536          & LSP   & 17.8-14.3  & 1.3-36.0  & 1.2x10$^{-7}$-2.7x10$^{-6}$ & 55575.29 - 56101.94    &...                 \\ %\hline
 {\bf 1ES~1426+428    }& 0.129       &  HSP      &   16.3-15.7  & 0.4-5.2  &... &   56047.00 - 56171.87                   & Fermi, KVA(pol) \\ %\hline
 {\bf PKS~1510-089  }  & 0.36         & LSP   & 16.6-13.1 &0.5-16.5 &2.6x10$^{-7}$-2.1x10$^{-5}$ &55575.30 - 56062.09    & ...                 \\ %\hline
 {\bf PG~1553+113  }   & \textless0.78  & HSP    & 14.0-13.1&   0.2-9.1 & 4.2x10$^{-8}$-1.0x10$^{-7}$  & 56007.13 - 56171.87    & ...                 \\ 
 {\bf Mrk~501    }& 0.034       & HSP    &    13.3-12.5  &0.8-6.6 & 3.9x10$^{-8}$-1.4x10$^{-7}$ &55660.04 - 56136.89                      & KVA (pol)          \\ %\hline
 {\bf BL~Lac      }& 0.069         & ISP    & 15.0-12.7      &  1.2-27.3 &10.0x10$^{-8}$-1.5x10$^{-6}$ & 55413.11 - 56225.94    & ...                \\ %\hline
\hline
\end{tabular}
}
\caption{The full RINGO2 catalogue with redshift, source type, R band magnitude range, Polarisation range, Fermi range, observation period information and details of absent/unavailable data (see Section \ref{fermi}). References for the redshift values can be found in Section \ref{description}.}\label{RINGO2_sample}
\end{minipage}
\end{table*}
%\end{threeparttable}

\subsection{Photometry}
\subsubsection{KVA 35cm}\label{KVA35}
The KVA telescope, operated remotely from Finland, consists of two tubes; 35cm and 60cm. The KVA 35\,cm is used for the R-band photometric observations of the Tuorla blazar monitoring program{\footnote{http://users.utu.fi/kani/1m}}. The observations are coordinated with the MAGIC Imaging Air Cherenkov Telescope and while the monitoring observations are typically performed two to three times a week (the weather allowing), during MAGIC observations the sources are observed every night. The data are analysed using standard aperture photometry procedures with the semi-automatic pipeline developed in Tuorla by K. Nilsson.  The pipeline presents the user with a graphical image of each frame to allow the rapid identification of the target object and comparison stars. The magnitudes are measured using the differential photometry and comparison star magnitudes found in the footnotes$^{1,2}$. The magnitudes are converted into Janskys using the standard formula $S=3080\times10^{-(\text{mag}/2.5)}$. For most of the sources, the contribution of the host galaxy to the measured flux is insignificant, the exceptions being Mrk~421 and Mrk~501. If the host galaxy has been detected, its contribution has been subtracted from the measured fluxes \citep{nilsson07}. Finally, the measured fluxes were corrected for the galactic absorption using the values from NED{\footnote{http://ned.ipac.caltech.edu}}.

While for many of the sources there is $>10$ years of data, in this paper, we only use the data that is from the same observing periods as our DIPOL and RINGO2 polarisation measurements.

\subsubsection{RINGO2}\label{RINGO2phot}
RINGO2 was a fast-readout imaging polarimeter with a V+R hybrid filter (covering 460-720 nm) constructed from a 3mm Schott GG475 filter cemented to a 2mm KG3 filter. RINGO2 used a rapidly rotating ($\sim$1 revolution per second) Polaroid to modulate the incoming beam from the telescope \citep{steele2010}. Eight equally spaced sensors around the edge of the rotating polaroid triggered the readout to a high speed electron-multiplying CCD. This  generated a series of frames (8 per rotation) of duration $\sim$125 msec each.  The frames were then averaged in software at matching rotation angles into a set of 8 images per observation. By this process the degree and angle of polarisation were encoded within the variation of the signal between the 8 images \citep{clarke&neumayer2002}. Data reduction was therefore a process of relative aperture photometry on these 8 images followed by the correction of the measured counts for instrumental polarisation and depolarisation based on the long term average properties of the nightly standard star observations.  

Sky-subtracted target counts were measured in each image using aperture photometry.  The associated error was computed by quadrature combination of the photon noise of the target, the sky noise in the aperture and error in sky determination. The photon noise was calculated according to an effective gain, that is, taking into consideration the multiple frames averaged to make a single image and the effect of multiplication noise in the electron multiplying charge coupled device (EMCCD) \citep{robbins2003}. The sky noise takes into account the number of pixels in the aperture and the sigma-cleaned standard deviation of sky annulus pixels. The sky error was calculated according to the number of pixels in the sky annulus.

RINGO2 produced 8 images, one for each of its Polaroid rotor positions. In order to obtain photometry measurements of these polarimetry data, the 8 rotor positions for a given observation were stacked using the IMCOMBINE command in IRAF. Automated relative aperture photometry was then performed on these frames using Source Extractor \citep{bertin1996} and the source was identified by locating the closest lying source to the right ascension and declination values. 

%The magnitudes were calibrated using the differential magnitude equation

%\begin{equation}
%m_1 = -2.5 . log_{10} \frac{F_1}{F_2} + m_2
%\end{equation}

%where m$_2$ is the apparent magnitude and F$_2$ is the flux of a non-variable secondary source and m$_1$ is the source apparent magnitude and F$_2$ is the source flux. 

The reference magnitudes for the secondary stars were found in the USNO-B catalogue \citep{monet2003} and where they were not available the magnitude value was offset at overlapping time periods to match the KVA data (which have a larger field of view so more choice of secondary stars). The magnitude was also converted into Janskys using the standard formula shown in Section \ref{KVA35}.

Full details of the RINGO2 and RINGO3 reduction pipeline can be found in \cite{jermak2016}.

%where F$_{Jy} $ is the flux in Janskys, 3080 is the conversion factor and m is the magnitude of the source. 
For Mrk~421 there were no usable secondary sources in the frame so magnitude calibration was not possible using the RINGO2 frames. SkyCamZ data were used instead.

\subsubsection{SkyCamZ}
The SkyCamZ camera consists of a 200mm diameter telescope that parallel points with the Liverpool Telescope in order to provide photometric monitoring during observations with other instruments and also carry out a synchronous variability survey of the northern sky.  The Z denotes a \textquoteleft zoomed\textquoteright field-of-view (1$^\circ$) and the instrument can detect sources down to $\sim$16 mag. When the enclosure is open, the camera takes a 10 second exposure automatically once per minute. All data are automatically dark subtracted, flat-fielded and fitted with a world co-ordinate system (WCS) by the STILT pipeline \citep{mawson2013}. The data are then introduced to the same pipeline used to reduce the RINGO2 data (see Section \ref{RINGO2phot}). The pipeline runs source extractor on the data and using a pre-identified secondary star (with its literature magnitude coming from the USNO-B1 catalogue) performs differential photometry.

\subsubsection{Fermi}\label{fermi}
Fermi-LAT (Large Area Telescope) is a space-based pair production telescope with an effective area of 6500cm$^2$ on axis for \textgreater 1 GeV photons). It is sensitive to $\gamma$-rays with energies in the range of 20 MeV to above 300 GeV \citep{atwood2009}. To produce the Fermi-LAT light curves the reprocessed Pass 7 data was downloaded and analysed using the ScienceTools version v9r32p5. In the event selection the LAT team recommendations were followed for Pass 7 data\footnote{http://fermi.gsfc.nasa.gov/ssc/data/analysis/documentation/Pass7REP\_usage.html}. We modelled a 15 degree region around each source using the instrument response function P7REP$\_$SOURCE$\_$V15, Galactic diffuse model gll$\_$iem$\_$v05$\_$rev1, and isotropic background model iso$\_$source$\_$v05.

The light curves were binned using an adaptive binning method \citep{lott2012}, with estimated 15\% statistical flux uncertainty in each bin. The flux in each bin was then estimated using the unbinned likelihood analysis and the tool gtlike. All sources within 15 degrees of the target that are listed in the 2FGL catalogue \citep{nolan2012} were included in the likelihood model. The spectral index of all sources are frozen to the values reported in 2FGL, and for sources more than 10 degrees from the target also fluxes are frozen to the 2FGL values. The sources Mrk~180 and 1ES1426+428 are too faint to produce adequate Fermi light curves for this analysis as the bin sizes would be too large.

\subsection{Optical Polarimetry}\label{OptPol}
\subsubsection{DIPOL at KVA-60}
The KVA polarisation monitoring program began in December 2008 using the Kungliga Vetenskapsakademien (KVA) telescope located on the Canary Island of La Palma. The KVA telescope consists of two telescopes; a 35~cm Celestron and a 60~cm Schmidt reflector. The larger of the two, DIPOL, a 60 cm reflector, is equipped with a CCD polarimeter capable of polarimetric measurements in BVRI bands using a plane-parallel calcite plate and a super-achromatic λ/2 retarder \citep{piirola2005}.

The observations typically took place 1-2 times a week. The typical observation time per source was 960s and the observations were performed without a filter to improve the signal-to-noise. There are several gaps in the cadence when the source has been too faint (R$>$15) and/or too weakly polarised (1-2\%) to be detectable with KVA. In total, 10 to $\sim$100 polarisation measurements per source were collected. During some of the nights, polarised standard stars from \cite{turnshek} were observed to determine the zero point of the position angle. The instrumental polarisation of the telescope has been found to be negligible.

The data analysis is performed following the standard aperture photometry procedures with the semi-automatic software that has been developed for monitoring purposes. The sky-subtracted target counts were measured in the ordinary and extraordinary beams using aperture photometry. The normalised Stokes parameters and the degree of polarisation and position angle were calculated from the intensity ratios of two beams using standard formula \citep[e.g.][]{pol}.

\subsubsection{RINGO2} \label{R2}
Optical polarimetry was obtained using the novel RINGO2 fast-readout imaging polarimeter \citep{steele2010} on the Liverpool Telescope (LT) (see Section \ref{RINGO2phot}). RINGO2 was mounted on the telescope in the period 2010 August 1 -- 2012 October 26. During this period observations were obtained of the blazar sample with a typical cadence of $\sim$3 nights whenever the source was observable from La Palma. Nightly observations were also obtained of polarised and non-polarised standards drawn from the catalogue of \cite{schmidt1992}.  Occasional more intensive (nightly) periods of observation were made of particular blazars when sources were in a high $\gamma$ ray state.   

The measured target counts were then corrected for instrumental polarisation by division by the corresponding mean value of the counts for the same Polaroid angle measured from all of the zero-polarised standard star observations (averaging over a period of time within which the polarimeter has not been removed from the telescope or altered). These corrected target counts and errors were then combined using the equations presented by \cite{clarke&neumayer2002} to calculate $q,u$ and their associated errors by standard error propagation. Analysis of the scatter in the $q,u$ polarisation values derived from the zero-polarised standards allowed us to estimate the stability of that correction as having an associated $q$ and $u$ errors of $0.25$\%, which we therefore combined in quadrature to our final error estimate. Next we combine $q$ and $u$ to estimate an initial value of degree of polarisation ($p$):

\begin{equation}
p = \sqrt{q^2+u^2}.
\end{equation}

This measured value was then corrected for an instrumental depolarisation factor 0.76$\pm$0.01 (D. Arnold, priv. comm.) which was found by plotting the measured polarised standards against their catalogue values. The depolarisation error is propagated into the degree of polarisation error. 

For PKS~1510-089 we used averaged data to account for the large scatter in the data points. The data were averaged over 5-day bins; within each bin we computed average q=Q/I and u=U/I values and corresponding root mean squared errors. 

\subsubsection{Polarimetric Error Analysis}
For both the DIPOL and RINGO2 polarimetric data, to correct $p$ for the statistical bias associated with calculating errors from square roots (where positive and negative $q$ and $u$ values are possible but only positive $p$ values can result) we used the methodology presented by \cite{simmons1985} to calculate 67\% confidence limits and the most likely $p$ value. As a check on this procedure we also ran a Monte Carlo simulation taking as input Gaussian distributions of $q$ and $u$ values with standard deviations equal to their calculated errors and examined the resulting distribution of $p$.  The results were identical.

The electric vector position angle (EVPA) in degrees was calculated as
\begin{equation}
{\rm EVPA} = {\rm atan2(u,q)} + {\rm ROTSKYANGLE} + {\rm PA}_0
\end{equation}

where the function atan2() calculates the arctangent of $u/q$ with a correct calculation of the sign and returns an angle between -180 and 180 degrees, ROTSKYANGLE is the angle of the telescope mount with respect to the sky when the image was taken and PA$_0$ is a calibration constant derived from repeated measurements of the EVPA of the polarised standard stars. Errors on EVPA were calculated according to the prescription in \cite{naghizadeh1993}, and again confirmed by Monte Carlo simulation.

\subsection{Identification of Flares and Rotations}
There are no exact definitions of what consists of a flare or a flaring period. By eye it is possible to identify datapoints that appear to be flaring however producing a sample wide condition that selects these points is difficult. It is not possible to assign a single level of quiescence above which data are considered to be flaring due to the varying baseline in $\gamma$-ray blazars. For clarity, we detail the conditions of the code used to identify the peak of $\gamma$-ray flaring periods. It is also necessary to identify what is considered to be an EVPA rotation or swing, this is detailed in this section and follows on from definitions in other EVPA studies. A summary of the results of this analysis is presented in Table \ref{rotstable}.

\subsubsection{Gamma-ray flares}\label{GRF}
A blazar flare may be associated with quasi-stationary, high density regions within the jet caused by magnetic field irregularities or it may be associated with a knot or blob of emission moving along the jet \citep{Astrophysics_processes}. The definition of a $\gamma$-ray flare is complicated due to it being relative to the (varying) baseline of the $\gamma$-ray emission at the time prior to the flare. The method used to identify flaring events was 2 fold. First we establish an initial level of increased activity using a moving window which defines \textquoteleft active\textquoteright\: points as those which are twice the standard deviation of the five preceding points. Then for the points that meet this criterion the condition of a flare is such; active points that are greater than five times the standard deviation of the preceding five active points are considered \textquoteleft flare\textquoteright. In addition, due to the nature of the moving window, for the first five points in the light curve if the flux is greater than the mean flux for the whole light curve then the points are classed as flares. Once the flare points are identified, a flare episode is defined by those flare points that are within 20 days of each other. These flaring episodes can often contain more than one peak, these are so close together that we define them as one event. The flaring episodes are represented in the light curves by vertical blue lines covering all flares within the 20 day range. For the analysis in Section \ref{evpaflare}, the {\it centre} of this flaring episode is used. %This method accounts for the changing quiescence level in these highly variable sources and attempts to identify the peak of a flaring episode, rather than the beginning and end of the flaring period.

%The following procedure was adopted: stepping along by 7 data points at a time, if the difference between the mean of these data points and the 8th datapoint is greater than 3 times the standard deviation of the 7 datapoints then the 8th point is identified as a flaring point. This means that those flares occurring at the start of the light curve (with no prior points) will not be identified, with the exception of the first $\gamma$-ray flare in 3C~66A. This process allows a non-biased flare identification process and identifies flaring periods which include the flares that occur before the baseline has dropped back to its previous value.

\subsubsection{Rotations of the polarisation angle}\label{pol_def}

The recurrent episodes of optical electric vector position angle (EVPA) rotations that are seen from AGN jets have been interpreted in a number of ways. Usually here we are referring to large-amplitude (\textgreater90$^\circ$), smooth and long-lasting rotation events which seem to signal some coherent process developing within the jet. Although random walks in the Stokes plane, driven by turbulent magnetic fields, have been demonstrated to be able to explain long rotations \citep{jones1985,marscher2014}, it cannot, for example, explain preferred rotation directions within some specific sources (which goes against the stochastic nature of the process). Nor (as shown via Monte Carlo simulations by \cite{blinov2015}) can they answer for an entire population of rotations observed. Other interpretations of the EVPA rotation which link them to coherent jet features, such as (a) plasma following a helical path due to a large-scale helicoidal magnetic field configuration of the jet, resulting in long, slow rotations of the EVPA \cite{marscher2008,marscher2010,zhang2014}; (b) a bend or curvature in the jet which leads to a projection effect on the plane of the sky akin to a rotation, which can invert its rotation due to relativistic effects resulting from the collimated emission \citep{abdo2010}.

Since the EVPA has a 180$^\circ$ ambiguity, long gaps in the polarisation light curves can lead to confusion when interpreting the EVPA rotations. To avoid introducing an incoherent view to the process with random 180$^\circ$ jumps being added to the EVPA dataset, we chose to interpret the observed EVPA light curves following a continuity hypothesis. We assume that variations proceed in the smoothest way possible with no sudden jumps. Although there is no predefined limit to the length of gaps in the data, we decide to apply correction only when the difference between consecutive data points is \textless30 days. In this work we define an EVPA rotation so that are results are consistent with those of the RoboPol group \citep{blinov2015}, therefore an EVPA rotation is \textquoteleft any continuous change of the EVPA curve with a total amplitude of $\Delta\theta_{max}$\textgreater90$^\circ$, which is comprised of at least four measurements with significant swings between them\textquoteright.

\section{Description of data}\label{description}
The flexibility of monitoring with a robotic telescope such as the Liverpool Telescope allows the user to increase monitoring of a particular source if its activity is deemed interesting. The main sample of blazars was added to over the period of the RINGO2 observations according to reported flaring and thus some sources have more seasons and more data than others. The multi-wavelength light curves of the individual sources in this sample can be found in the Appendix.  

\subsection{3C~66A}
3C~66A is a well-known BL Lac at redshift z$>= $0.3347 \citep{furniss2013}. The blazar is a bright source of HE (Acciari et al. 2009) and very high energy (VHE) $\gamma$-rays (E \textgreater 100 GeV)  \citep{2009ApJ...692L..29A, acciari2009}. There are many polarimetric monitoring observations of this source (e.g. \cite{takalo1993} and references therein). In these data the polarisation degree is always high, typically between 10 - 20 \% with the maximum value measured 33 \%. In the historical data the EVPA is significantly variable but shows a preferred position angle around 20 - 40$^\circ$, which is perpendicular to the direction of the VLBA jet. \cite{ikejiri2011} and \cite{itoh2013} also report a rotation of EVPA of \textgreater180$^\circ$ (at MJD$\sim$54840, early January 2014) and a significant negative correlation of the flux and polarisation degree. \cite{itoh2013} separate their $\sim$2 years worth of data into four sections according to the optical flux and polarisation degree. Our program does not cover their first and second periods.

During the first season our polarisation data show similar behaviour to that in Itoh in the period spanning August to November 2009 (MJD 55048-55150). We see a $\sim$180$^\circ$ rotation of the EVPA, however, during November 2009 to January 2010 (MJD 55151 to 55220) our data appear to show another 180$^\circ$ rotation (see Table \ref{rotstable} and Figure \ref{fig:3c66a} in the Appendix). The nature of the $\pm$180$^\circ$ ambiguity and the smooth rotation selection of the EVPA data means that the absence of even one data point can be the difference between a rotation (our data) or a slight peak (Figure 4 in \cite{itoh2013}). However we have combined our data with that of Itoh (priv. comm.) and we see the combined data suggest a rotation. \cite{itoh2013} describe a polarisation degree which is systematically different among the four periods due to a long-term slow change. We continue to see this behaviour in our data beyond their fourth period. The source enters a relatively quiescent phase after late July 2010 (MJD$\sim$55400) and in the remaining three epochs the EVPA does not appear to rotate. The optical flux drops to $\sim$14.5 in the R band and the $\gamma$-ray flux stays low with the exception of a small flare at MJD$\sim$56150 (August 2012).

\subsection{S5~0716+714}
The BL~Lac object S5~0716+714 has been studied intensively at all frequencies. It has no spectroscopic redshift but constraints from intervening absorption systems give z\textless0.322 (95\% confidence level, \cite{danforth2013}) and the host galaxy detection z=0.31$\pm$0.08 \citep{nilsson07}. S5 0716+714 is a bright source of HE \citep{acciari2009} and VHE \citep{anderhub2009} $\gamma$-rays. The source is thought to be observed very close to the line of sight of the jet allowing an excellent view down the jet itself \citep{impey2000}. There are several dedicated studies of the optical polarisation behaviour of the source \citep{uemura2010}. In the optical band the source shows extremely fast brightness and degree of polarisation variations. Intra-night variability of the polarisation has been reported by \cite{impey2000}and \cite{villforth2009} with significant variations on timescales of 10-15 minutes. In our data we also see fast brightness and degree of polarisation variations across the four seasons with a variation of $\sim$2 magnitudes in the optical and a range in degree of polarisation of 0-$\sim$20\%.

Typically the source shows no correlation between optical brightness and polarisation degree \citep{ikejiri2011} and in this work we too find no evidence of a correlation (see Table \ref{full_table}). S5~0716+714 also shows no apparent long-term trends and therefore it has been suggested that at all epochs there must be a number of polarisation components showing variations on a small timescale. The literature also reports rapid (6 days) \citep{larionov2008a} and slow (400 days) rotations of the EVPA of \textgreater300$^\circ$ \citep{ikejiri2011}. In the historical data the range of the degree of polarisation is from $\sim$0-25\%. The interpretation of the $\pm$180$^\circ$ ambiguity in our data indicates either the EVPA exhibits a rapid rotation of $\sim$180$^\circ$ in March 2009 (MJD$\sim$54900), a long slow rotation of $\gtrsim$200$^\circ$ in October 2010 (MJD$\sim$55550) and a rapid rotation of $\sim$200$^\circ$ in March 2012 (MJD$\sim$56000). A rotation which changes direction mid rotation, such as the rotation in March 2012 (MJD$\sim$56000) may be due to a knot of material crossing the observers line of sight (see \cite{aleksic20143c2} with reference to 3C279) however, this behaviour may also be due to a random walk of the EVPA. The $\gamma$-ray and optical data show correlations in late 2009 and early 2011, see the Discussion for more details.

\subsection{OJ~287}
OJ~287 (z=0.305) is a BL~Lac object and one of the most famous blazars as it hosts a supermassive {\it binary} black hole system at its centre \citep{sillanpaa1988}. It is bright in HE $\gamma$-rays \citep{acciari2009} but has not been detected in VHE $\gamma$-rays (e.g. \citep{2009PASJ...61.1011S}). The dedicated studies of the optical polarisation behaviour \citep{2009ApJ...697..985D, 2010MNRAS.402.2087V, uemura2010} have shown that there is a strong preferred position angle for the polarisation which is perpendicular to the flow of the jet. The polarisation is strong (maximum 35\%). Occasionally the EVPA also shows rapid rotations with durations of 10-25 days. This behaviour has been interpreted as a signature of two components \citep{1984MNRAS.211..497H, 2010MNRAS.402.2087V}, stationary polarisation core and chaotic jet emission. Occasionally flares with a negative correlation between flux and polarisation degree have been observed \citep{ikejiri2011}. Our data include the period observed by \cite{agudo2011} and we see similar behaviour of the polarisation properties, particularly the rotation in April 2009 (MJD$\sim$54940) which displays peculiar behaviour.

 With the exception of the rotation in April 2009, the EVPA remains relatively stable throughout the next three observing seasons. The degree of polarisation is high in the first and fourth seasons but drops from $\sim$30\% to \textless5\% during the second season following a $\gamma$-ray flare. The optical flux is variable and ranges between 13.5 and 15.5 magnitudes.

\subsection{1ES~1011+496}
1ES~1011+496 (z=0.212, \cite{albert2007b}) is a BL Lac object which, until its discovery in VHE $\gamma$-rays \cite{albert2007a}, was little studied. It is bright in HE $\gamma$-rays \citep{acciari2009} and has little optical polarisation literature data. From 1987 there is one archival polarisation observation \citep{wills2011} which shows low polarisation $\sim$2\% and an EVPA $\sim$86$^\circ$. The KVA and RINGO2 data presented here (taken for multi-wavelength campaigns \citep{ahnen2016a,ahnen2016b} show similar results to the archival observations. Low polarisation of \textless10\% and an EVPA at $\sim$110$^\circ$.

\subsection{Mrk~421}
Mrk~421 (z=0.03 \citep{devaucouleurs1991} is a nearby BL Lac object that was the first extragalactic VHE $\gamma$-ray emitter to be discovered \citep{punch1992}. Its optical polarisation behaviour has been studied extensively in the past (e.g. \cite{hagentorn1983,tosti1998b} and references therein). In these data for the majority of the time the source shows rather low polarisation \textless5\% and a rather stable EVPA of $\sim$180$^\circ$. \cite{ikejiri2011} found overall significant correlation between optical brightness and degree of polarisation and during a large optical flare in the winter of 1996-1997 (MJD$\sim$50350 onwards) the optical degree of polarisation rose to 12\% \citep{tosti1998b}. \cite{tosti1998b} reported a rotation of the position angle of $\sim$90$^\circ$ from May to October 1995 (MJD$\sim$49838-49991).

The data collected with the KVA and RINGO2 from 2008-2011 (MJD 55409-56226) agree with this general behaviour, however from the beginning of 2012 (MJD$\sim$55900 onwards) the data show a strong increase in polarisation degree and a 360$^\circ$ rotation of the position angle, along with a steady increased in the optical magnitude. This behaviour precedes an unprecedentedly large $\gamma$-ray flare which occurs after June 2012 (MJD$\sim$56100) and unfortunately is not accompanied by optical data due to the source's visibility.

\subsection{Mrk~180}
Mrk~180 (z=0.045 \citep{falco1999}) is a nearby BL Lac object that was detected in VHE $\gamma$-rays in 2006 (MJD$\sim$53795) \citep{albert2006}. The source has a bright host galaxy and is little studied in optical polarimetry. \cite{marcha1996} presented an optical degree of polarisation measurement of 2.4\% which is consistent with the results we see from early 2012 (MJD$\sim$56000) when our observations begin. The RINGO2 data were taken as part of a multi-wavelength campaign which was started in 2008 (MJD$\sim$54500) \citep{rugamer2011}. We have few data points for the RINGO2 period of observation, polarimetry from the KVA is unavailable as are Fermi flux data. Our monitoring continues with the third generation polarimeter RINGO3 \citep{arnold2012}. We find a stable EVPA of $\sim$40 $^\circ$ and an R band magnitude of $\sim$14.4.

\subsection{1ES~1218+304}
1ES~1218+304 is a BL Lac source at z=0.162 \citep{adelman2009}. It is a source of HE and VHE \citep{albert20061218} $\gamma$-rays. Its optical flux has varied over the last $\sim$10 years from 15.2-16.4 magnitudes in the R band (from Tuorla blazar monitoring campaign$^1$). There are very few polarisation measurements in the literature, \cite{jannuzi1994} report a degree of polarisation of $\sim$5\%. We present KVA (photometry only) and RINGO2 (polarimetry and photometry) data from March 2012 (MJD$\sim$55987) until the end of the program, the source does not show any flares or rotations. 

\subsection{ON~231}
ON~231, also commonly known as W Comae, (z=0.102 \cite{weistrop1985}) is a HE and VHE $\gamma$-ray bright BL Lac object \citep{acciari2008,acciari2009}.  Observations taken in 1981-1982 (MJD$\sim$45000) and 1991-1992 (MJD$\sim$48500) cover similar ranges with a polarisation degree ranging from 5.2-19.2\% and an EVPA of 50.0-93.3$^\circ$ \citep{wills2011}. The source underwent three major outbursts in March 1995 (MJD$\sim$49800), February 1996 (MJD$\sim$50120) and January 1997 (MJD$\sim$50450) \citep{tosti1998a} and a more energetic outburst in June 1998 (MJD$\sim$50800-51000) which saw the brightness increase by $\sim$3 magnitudes in the R band \citep{tosti1999}. \cite{ikejiri2011} observed this source from 2008-2010 and during this period the polarisation was 3.5-19.6\% with an EVPA of 60-80$^\circ$. They also found significant negative correlation between flux and polarisation degree (see Figure 28 of \cite{ikejiri2011}). \cite{sorcia2014} presented results from February 2008 to May 2013 and find a gradual decrease in mean flux over the $\sim$5 year period of $\sim$3~mJy. They find a maximum degree of polarisation of 33.8\% $\pm$ 1.6\% and a large rotation of the EVPA of $\sim$237$^\circ$ which coincided with a $\gamma$-ray flare in June 2008. 

The KVA-60 and RINGO2 data in this work show a degree of polarisation and EVPA consistent with the source in a low state. We see slightly brighter optical and $\gamma$-ray fluxes in the first season (see Figure \ref{fig:on231} in the Appendix) and optical magnitude starts to decrease with the increase in degree of polarisation at the end of our last observing season.

\subsection{PKS~1222+216}
PKS~1222+216 (z=0.435, \cite{veroncetty2006}) is a flat spectrum radio quasar (FSRQ), and therefore LSP, which has received a lot of attention since its discovery in VHE $\gamma$-rays \citep{aleksic2011a}. Very little optical polarisation data are available in the literature. A single measurement from \citep{ikejiri2011} shows a degree of polarisation of 5.9 \%. The data presented here were taken in the 2011-2012 (MJD$\sim$55600-56100) observing season when the source was in a quiescent state in optical and $\gamma$-rays. The EVPA shows very little variation and the degree of polarisation is low (\textless10\%).

\subsection{3C~279}
3C~279 (z=0.536, \cite{burbidge1965}) was one of the first extragalactic $\gamma$-ray sources discovered \citep{1992ApJ...385L...1H} and is one of the first flat spectrum radio quasars to be detected in very high energy $\gamma$-rays \citep{albert2008}. Over $\sim$10 years of observations the source showed variability ranging from 13-16 magnitudes in the R band. In the space of $\sim$100 days the source became fainter by 3 magnitudes \citep{larionov2008b} and from MJD~54120-54200 showed a rotation which they conclude is intrinsic to the jet. This rotation was coincident with a low degree of polarisation which was higher before and after the rotation (at 23\%). The low polarisation during the rotation is attributed by the authors to the symmetry of the toroidal component of the helical magnetic field. In the period prior to the the start of the RINGO2 program the source showed a rapid decline in magnitude over the period of $\sim$1 month synchronous with an increase in polarisation degree and a 180$^\circ$ rotation of the position angle (\citep{abdo2010}. \cite{kiehlmann2013}, using data from RINGO2 and KVA-60 amongst other instruments, showed there was an increase in flux and degree of polarisation along with a $\sim$150$^\circ$ rotation of the position angle in May 2011 (MJD 55700) the addition of Fermi data showed that during this period of $\sim$2 months the $\gamma$-ray flux decreases by $\sim$100 [10$^{-8}$ cm$^{-2}$ s$^{-1}$] \citep{aleksic20143c2}. \cite{aleksic20143c2} interpreted this optical outburst with a rotation of the position angle and the increase in the degree of polarisation as geometric and relativistic aberration effects such as an emission knot's trajectory bending such that it crosses the observer's line of sight (for full description see \cite{aleksic20143c2}). 

We have three seasons of polarimetric data and four seasons of photometric data. Having the same data as \cite{kiehlmann2013} we see the same behaviour. In the third and fourth seasons we see rotations that rotate in one direction and then back on themselves. We see an additional rotation which is followed by a lack of data. The source drops in brightness at the start of the observing period and is at its highest in polarisation ($\sim$40\%). The source is brightest in the optical in the last two seasons of observations where the degree of polarisation dropping to lower values.

\subsection{1ES1426+428}
1ES1426+428 is a lesser-studied BL Lac object at a redshift of z= 0.132 \citep{urry2000}. It is classified as an extreme high-energy-peaked BL Lac object (HSP) \citep{costamante2001} and is a HE $\gamma$-ray source \citep{petry2000, horan2002}. \cite{jannuzi1994} report an optical degree of polarisation of \textless7 \% over the period of 1988-1990 MJD$\sim$47161 - 47892) and a non-constant EVPA which may show a slow rotation of $\sim$150$^\circ$. The optical flux maintains a fairly constant value of between 16-17 in the R band. We find the degree of polarisation stays below $\sim$6\%, the optical flux is constant with a median R band magnitude of 15.6. The EVPA stays relatively constant within the error bars. The source is too faint in $\gamma$-rays to be significantly detected within the analysed time window.

\subsection{PKS~1510-089}
PKS~1510-089 is a $\gamma$-ray bright \citep{acciari2009} FSRQ/LSP at z=0.36 \citep{burbidge1966}. The source has shown bright flares in optical, radio, x-ray and HE $\gamma$-rays at the beginning of 2009 (MJD$\sim$54910-54920), along with the first detection of VHE $\gamma$-rays \citep{2013AandA...554A.107H}. During the $\gamma$-ray flaring from MJD~54950~-~MJD~55000 (April 2009 onwards) the optical electric vector position angle (EVPA) was reported to rotate by \textgreater 720$^\circ$ and during the major optical flare the optical polarisation degree increased to \textless30\% \citep{marscher2010}. In early 2012 (MJD 55960 and onwards) it again showed high activity in HE $\gamma$-rays and was also detected in VHE $\gamma$-rays. Again there was a \textgreater 180$^\circ$ rotation of the EVPA following this flare but the polarisation degree stayed low $\sim$2\% \citep{aleksic2014}.

\begin{figure}
\centering
\includegraphics[width=8cm]{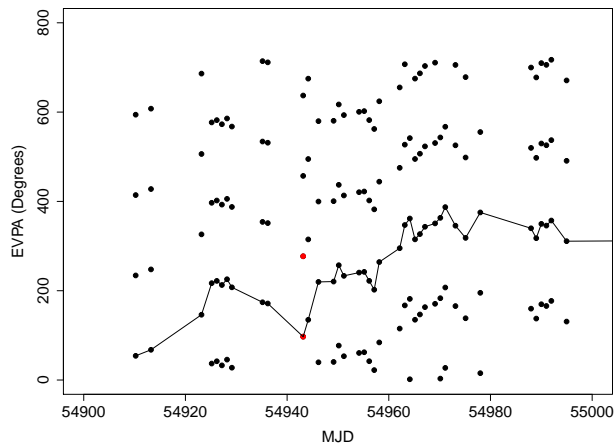}
\caption[]{Zoomed view of PKS~1510-089 light curve (full light curve shown in Appendix, Figure \ref{fig:pks1510}). This plot can be compared with Figure 4 in \cite{marscher2010} where the polarisation angle data are interpreted as showing a 720$^\circ$ rotation. Here we report a rotation of less than half that (333$^\circ$) and this is due to the interpretation of this particular dataset by the EVPA trace code which, to account for the $\pm$180, ambiguity identifies the next nearest lying point. The points in red highlight where the difference in identification originates.}
\label{fig:pks1510_zoom}
\end{figure}

The polarimetric RINGO2 data we present for this source has been averaged over 5-day bins to account for the scatter in the data (see Section \ref{R2}). Our data show the above mentioned $\gamma$-ray and optical flaring activity from the end of 2008 into 2009, we see a rotation in the EVPA at this time but due to interpretation of our data using the EVPA tracing code (see Section \ref{pol_def}) we do not report a rotation as great as 720$^\circ$, rather a rotation of 333$^\circ$. The difference is due to the data sampling and thus highlights the need for intensive optical monitoring during $\gamma$-ray activity. For clarity we include a zoomed region of Figure \ref{fig:pks1510} (see Figure \ref{fig:pks1510_zoom}) for comparison with the bottom panel of Figure 4 in \cite{marscher2010}. The red points show the EVPA data point at it is measured and at the +180 position. The EVPA trace code in this work selected the lower of the two points as it is closer to the previous point, had there been intermediate points the rotation might have shown to continue at a steeper gradient which would result in a $\sim$600$^\circ$ rotation measurement.

\cite{ikejiri2011} report a correlation between V band magnitude and degree of polarisation. We see similar results in our analysis (see Table \ref{full_table}), however, because we lack data when the R band magnitude was the brightest we are unable to populate the brighter end of the magnitude-degree of polarisation plot.

%Our data appear to show different behaviour to that reported in the previous two papers, however, on closer inspection it seems that again the difference is due to sampling and highlights the need for intensive optical monitoring when a source flares in $\gamma$-rays. \cite{ikejiri2011} report a correlation between V band magnitude and degree of polarisation. We see similar results in our analysis, however, because we lack data when the R band magnitude was the brightest we are unable to populate the brighter end of the magnitude-degree of polarisation plot.

\subsection{PG~1553+113}
PG~1553+113 is a $\gamma$-ray bright BL Lac object at z\textgreater0.395 \citep{danforth2010}. It is a persistent source of VHE $\gamma$-rays \citep{albert2007a,aharonian2006} which has triggered several multi-wavelength studies of the source (e.g. \cite{aleksic2010}). However, only few campaigns have included polarimetric observations. Polarisation observations were reported in \cite{albert2007a}, \cite{andruchow2011} and \cite{ikejiri2011} with the maximum value for polarisation degree of 8.2\%. The observations of \cite{ikejiri2011}, which cover the longest period of time, do not show a clear preferred angle for the EVPA. In 2008 (MJD$\sim$54600 onwards) the EVPA was $\sim$100-150$^\circ$ while the later observations (in 2009 and 2010 (MJD$\sim$55100 and 55200) only single data points) show an EVPA $\sim$50$^\circ$. RINGO2 and KVA-60 data suffer from poor sampling but agree with literature, showing an EVPA which is $\sim$100-150$^\circ$ until March 2012 when there is a rotation over a period of a few months which coincides with a flare detected in HESS and MAGIC but not in Fermi \citep{abramowski2015,aleksic2015}.

\subsection{Mrk~501}
Mrk~501 (z=0.0337, \cite{ulrich1975}) is a BL Lac type source which was discovered as a VHE $\gamma$-ray source in 1996 \citep{quinn1996} and above 1.5 TeV \citep{bradbury1997}. The source was observed during a period of high activity in 1997 (MJD$\sim$50449) with a degree of polarisation of 1-3\% \citep{joshi2000} and R band magnitude $\sim$13.5 \citep{petry2000}. The source showed a VHE flare in 2009 (MJD$\sim$54940) which was correlated with a 5\% increase in the optical degree of polarisation, a significant increase compared to the typical polarisation level of 1-3\% \citep{ulisses2011}. Along with the increase in degree of polarisation, the EVPA rotated by 15$^\circ$. In the available data we see no rotations of the EVPA (see definition of EVPA in Section \ref{pol_def}) and a very stable optical flux, the degree of polarisation reaches $\sim$6\% which could be correlated with a small $\gamma$-ray flare, however, for the larger $\gamma$-ray flare the source was not visible from La Palma.

\subsection{BL~Lac}
BL~Lac (z=0.069, \cite{vermeulen1995}) is a bright source of HE $\gamma$-rays and occasionally of VHE $\gamma$-rays \citep{albert2007b,Arlen2013}. Its polarisation has been extensively studied with the two long-term studies presented in \cite{hagenthorn2002a} and \cite{hagenthorn2002b}. In these publications observations from 1969 to 1991 (MJD$\sim$40000-48500) were presented. It was found that the EVPA showed a preferred angle of $\sim$22$^\circ$ which is close to the direction of the jet in very long baseline interferometry (VLBI). In the second half of their data (1980-1991, MJD$\sim$44239-48500) BL Lac showed significant periodicity of 308 days both in total flux and relative Stokes parameter $q$ \citep{hagenthorn2002a}. The polarisation degree for this 22 year period varied from \textless1\% to $\sim$40\%.

In October 2005 (MJD$\sim$53660) the source underwent a double peaked optical outburst with rapid rotation of the EVPA during the first outburst, this led \cite{marscher2008} to conclude that the rotation took place in the collimation and acceleration zone of the jet where a helical magnetic field would be present. In June 2011 (MJD$\sim$55710) a rapid TeV flare was detected which coincided with a rapid change in optical polarisation angle \citep{Arlen2013}. This was concluded to support the model of \cite{marscher2008}. \cite{raiteri2013} whose data coincide with that of this work, report an EVPA of $\sim$15$^\circ$ which is nearly aligned with the radio core EVPA and mean jet direction. 

The EVPA tracing code presented in this paper (see Section \ref{pol_def}) identifies four polarisation angle rotations in the BL~Lac data, however, only two of these can be classified as \textquoteleft true \textquoteright rotations according to the condition that the rotation must consist of 4 or more measurements with significant swings between them. The degree of polarisation varies between values of $\sim$25\% and little or no polarisation signal at all. The drop to a degree of polarisation of $\sim$0 coincides with a EVPA rotation and a $\gamma$-ray flare, along with an increase in the optical magnitude, and is consistent with previously reported behaviour.

\section{Discussion}\label{disc}
In this section comparisons are made between the polarisation properties, optical flux and $\gamma$-ray flux for those sources with reasonable sampling. This sample is subject to selection biases and therefore the results in this work cannot be generalised to the larger blazar population. For those sources which have been observed only for a short period of time, which have sparse data sampling or lack sufficient multi-wavelength information  (Mrk~180, 1ES~1011+496, 1ES~1426+496 and 1ES~1218+304) only optical-optical analysis and their light curves (see Appendix) are presented and they are excluded from the $\gamma$-ray analysis. The following section explores correlations between the optical data and the $\gamma$-ray flux, along with the frequency of flares in relation to optical polarisation rotations. 

\begin{table*}
\centering
\resizebox{\textwidth}{!}{
\begin{tabular}{|l|l|l|l|l|l|l|l|l|l|}
\hline
{\bf 1. Source}& \textbf{2. Rot $\uparrow$} & \textbf{3. Rot $\downarrow$} & \textbf{4. Flares} & \textbf{5. Type} & \textbf{6. Fermi mon.} & \textbf{7. Max.} & \textbf{8. Flare rate} &\textbf{9. Days between }&\textbf{10. Flares}\\ 
                 & \textbf{ (anti-c-wise)} & \textbf{(c-wise)} && &{\bf period (days)}&\textbf{deg.}&\textbf{flares/day (year)}&{\bf rot \& flare}&{\bf  during rot}\\ \hline
\textbf{3C~66A}   & 1                              & 0             & 10           & ISP           & 1323            & 27.7              & 0.0076 (2.76) 	&20 			& 0    \\ 
\textbf{S5~0716+714}  & 2                      & 2             & 16              & ISP           & 1442            & 17.9              & 0.011 (4.05)  	&-14, 0 (x3) 			&  3       \\ 
\textbf{OJ~287}   & 1                               & 0             & 3              & LSP           & 1382            & 36.9              & 0.0022 (0.79)  	&...		 			& ...	 	 \\ 
\textbf{Mrk~421}  & 0                              & 1              & 3             & HSP           & 541             & 8.3               & 0.0055 (2.03)	&1 					&0  \\ 
\textbf{ON231}   & 0                                & 0             & 5             & ISP           & 1593            & 23.3              & 0.0031 (1.15)   	&...					&... 		 \\ 
\textbf{PKS~1222+216} & 0                     & 0              & 5               & LSP (FSRQ)        & 530             & 5.4     & 0.0094 (3.45)  	&...					& ...        	\\
\textbf{3C~279}   & 3                              & 1             & 11           & LSP (FSRQ) & 1464           & 29.7              & 0.0075(2.74)  	&8, -2, 0 				 & 1   \\ 
\textbf{PKS~1510-089} & 3                      & 1             & 24         & LSP (FSRQ)  & 1406            & 16.4              & 0.017 (6.23)	 &-4, 21, 39, -84, 0 (x10) 	 \\ 
\textbf{PG~1553+113}  & 1                     & 1              & 5              & HSP           & 1241            & 9.1               & 0.0040 (1.47)     &0			&  1 	\\ 
\textbf{Mrk~501}  & 0                              & 0              & 4              & HSP           & 594             & 6                 & 0.0067 (2.46)     &...					& ...		\\ 
\textbf{BL~Lac}  & 2                                 & 0             & 9             & ISP           & 1431            & 27.3             & 0.0063 (2.30) 	&116, -21, 0, 98 		&  1   \\ 
\hline
\end{tabular}}
\caption{Tabulated data of the upward and downward EVPA rotations and $\gamma$-ray flares for different blazar subclasses for the 11 sources that have EVPA rotation/$\gamma$-ray flare events. Also included are the length of the Fermi monitoring period in days, the maximum degree of polarisation, the flare rate (and mean flare rate), days between rotations and flares (and mean of this value) and number of flares during a rotation.}
\label{flare_rot_data}\label{rotstable}
\end{table*}

\subsection{Correlation analysis}\label{correlationanalysis}
Correlations between the optical and $\gamma$-ray data can give information about the emission regions and magnetic field structure within the jets of the different blazars. Optical flux (lacking a strong polarisation signal) can also originate from outside the jet. 

Due to observational constraints from ground-based telescopes, along with weather and observing priorities, we have imperfectly sampled optical data. While it is possible that the continuous Fermi $\gamma$-ray data could be binned to coincide with optical monitoring this could not be possible with the adaptive binning code used in this work as it automatically sets the bin sizes according to the gamma-ray brightness of the source (see Section \ref{fermi}).  The binning of $\gamma$-ray data according to optical observations is likely to dilute flaring behaviour (which is displayed in more detail with the adaptive binning method) and also involve difficulties in establishing bin sizes because the optical observations only take $\sim$1-5 minutes.

In order to match the optical and $\gamma$-ray data points (which are of course not completely synchronous) we explore two methods. In method one we use each of the dates associated with optical observations and interpolate a value from the $\gamma$-ray light curve for this date by fitting a gradient to the nearest neighbouring $\gamma$-ray points and calculating the matched $\gamma$-ray flux using the equation for a straight line. The plots in Figure \ref{fig:mag_gam_plot} are produced by such a method and show the overall behaviour of the sources according to their different subclasses.  

It is also possible to match the optical and $\gamma$-ray data by using the same bins as the $\gamma$-ray data to bin the optical data. This method produces correlation plots (see Figure \ref{fig:GB_mag_gam_plot}) which are less dense than those produced by leading with optical data sampling (compare with Figure \ref{fig:mag_gam_plot}). Binning of optical data in this way results in higher temporal frequency optical activity being averaged out,  As the focus of this paper is the optical data we therefore use our first method in the following analysis. The same method is applied using the optical polarisation degree dates. When data from one wavelength do not change over a period in which data from the other wavelength does change then these periods appear on the correlation plots as straight horizontal or vertical lines.

We use the Spearman Rank Coefficient test to determine the correlation of the data. The null hypothesis states that the two variables are not correlated. If $p<0.05$ then the null hypothesis can be rejected.  Significant correlations are indicated by $p>0.95$ (no correlation) or $p<0.05$ (correlation).  For the analysis that involves magnitude, the values of the correlation coefficient $\rho$ have been calculated so that the reverse nature of the parameter is appropriately used.

For exploring the distributions of Spearman Rank test results, we will use the Kolmogorov~Smirnoff (KS) test where the null hypothesis is that the two samples are drawn from the same population where p=1 suggests there is a strong probability that the samples come from the same parent distribution.

%The optical photometric and polarimetric data are obtained simultaneously to each other by the LT and KVA separately. With the focus of this paper being the optical properties of the sources we lead with our optical data and use the modified julian data (MJD=JD-2400000.5) (to 5 significant figures) to interpolate from the two nearest $\gamma$-ray data points. This results in synchronous estimates of the Fermi flux which we plot against the available optical data points. The same method is applied using the optical polarisation degree dates. When data from one wavelength source do not change over a period in which data from the other wavelength source does change then these periods appear on the correlation plots as straight horizontal or vertical lines.

%\begin{table}
%\centering
%\label{corrtable}
%\begin{tabular}{llll}
%\hline
 %               & \textbf{$\rho$} & \textbf{p} & \textbf{N} \\ \hline
%\textbf{All}    & 0.257           & 1.000          & 13         \\
%\textbf{FSRQ}   & 0.066           & 0.875      & 3          \\
%\textbf{BL Lac} & 0.703           & 1.000          & 10         \\
%\textbf{LSP}    & -0.527          & 1.000          & 4          \\
%\textbf{ISP}    & 0.644           & 1.000          & 4          \\
%\textbf{HSP}    & 0.793           & 1.000          & 5      \\ \hline  
%\end{tabular}
%\caption{Spearman rank correlation coefficient $\rho$, p value and number of sources for each blazar subclass for the $\gamma$-ray and optical flux correlations (See Figure %\ref{fig:mag_gam_plot} for plots).}
%\end{table}

\begin{table}
\centering
\label{corrtable}
\begin{tabular}{llll}
\hline
                & \textbf{$\rho$} & \textbf{p} & \textbf{N} \\ \hline
\textbf{All}    	& -0.153          & 6.4x10$^{-14}$         			& 13         \\
\textbf{FSRQ}   & 0.149           & 7.3x10$^{-4}$      			& 3          \\
\textbf{BL Lac} & 0.395           & \textless2.2x10$^{-16}$          	& 10         \\
\textbf{LSP}    	& -0.539          & \textless2.2x10$^{-16}$          	& 4          \\
\textbf{ISP}    	& 0.578        	& \textless2.2x10$^{-16}$     		& 4          \\
\textbf{HSP}    	& 0.597           	& \textless2.2x10$^{-16}$          	& 5      \\ \hline  
\end{tabular}
\caption{Spearman rank correlation coefficient $\rho$, p value and number of sources for each blazar subclass for the $\gamma$-ray and optical flux correlations (See Figure \ref{fig:mag_gam_plot} for plots).}
\end{table}

\subsubsection{Optical and $\gamma$-ray flux correlations}\label{opt_gam_0}
\begin{figure*}
\centering
\includegraphics[width=18cm]{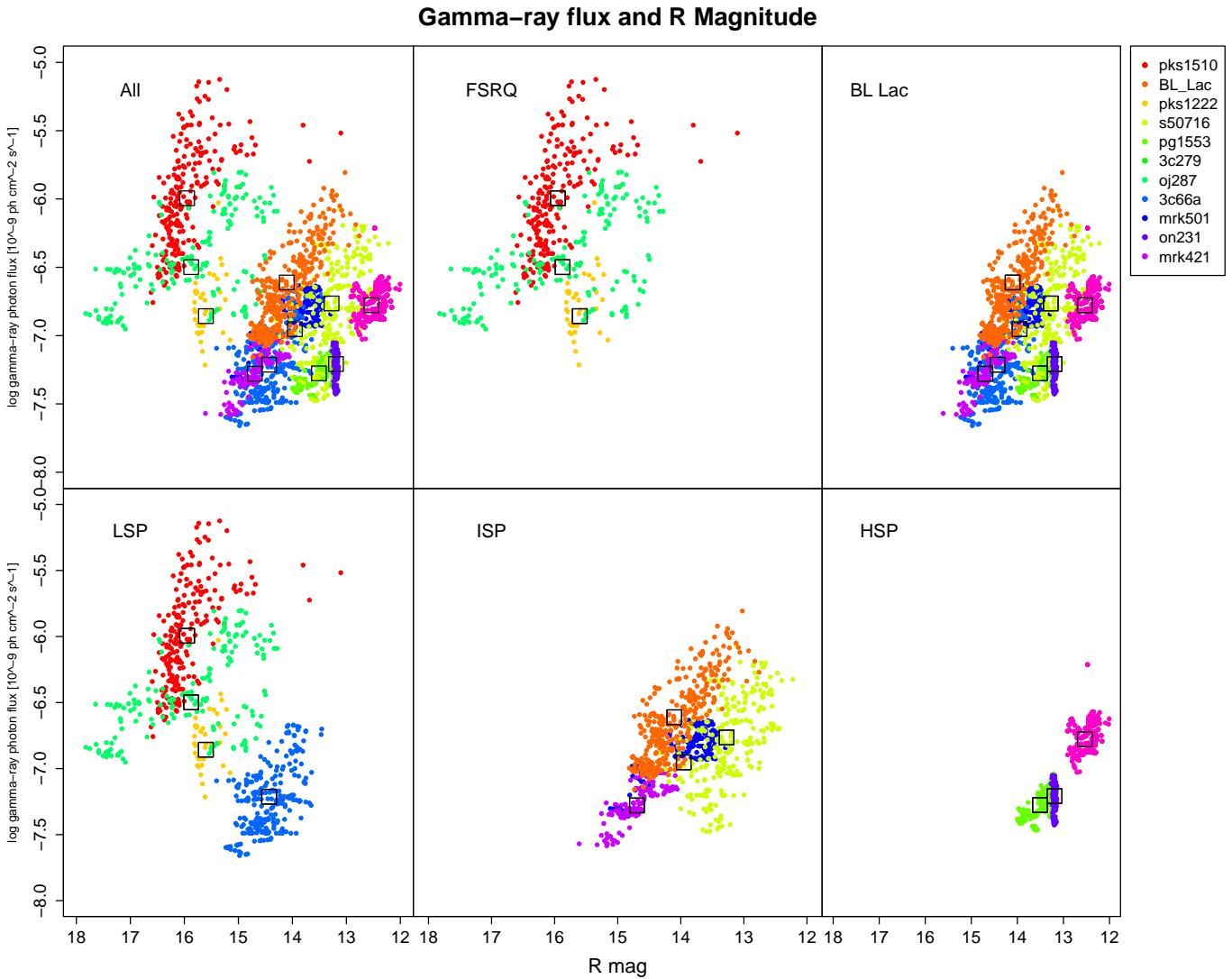}
\caption[]{Fermi $\gamma$-ray data plotted against magnitude for 11/15 sources (those which have \textgreater5 $\gamma$-ray datapoints) (each with a separate colour) and subsequent blazar subclasses: FSRQs, BL~Lacs, LSPs, ISPs and HSPs. The $\gamma$-ray data points are interpolated to match the date of the optical data points (see Section \ref{correlationanalysis}). Black squares show the mean $\gamma$-ray and optical value for each source.}
\label{fig:mag_gam_plot}
\end{figure*}

\begin{figure*}
\centering
\includegraphics[width=18cm]{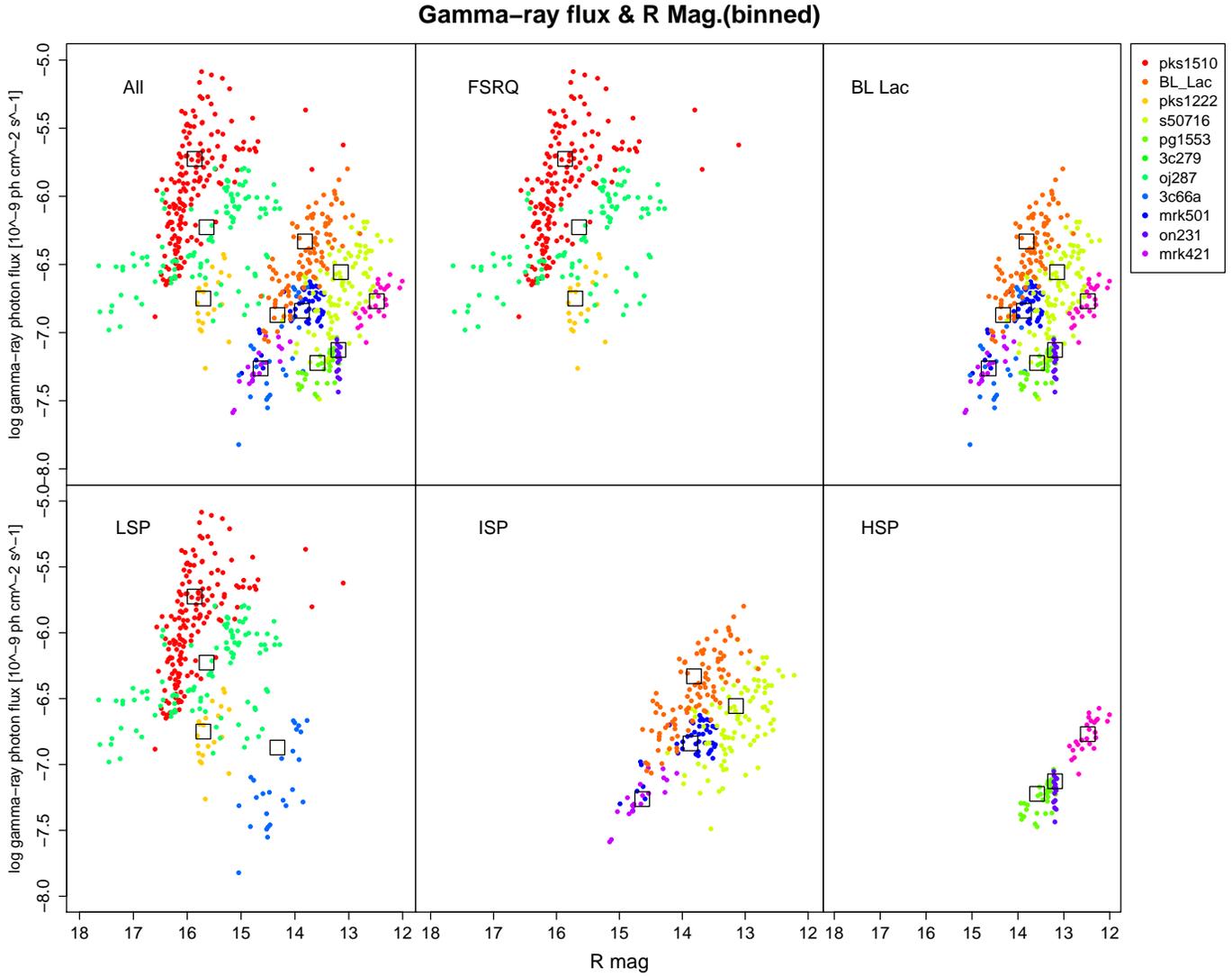}
\caption[]{The sample plots as Figure \ref{fig:mag_gam_plot} with optical data points binned according to the range of the Fermi bins. There are fewer data but the overall trends are similar.}
\label{fig:GB_mag_gam_plot}
\end{figure*}

Figure \ref{fig:mag_gam_plot} shows 6 plots of $\gamma$-ray flux against optical magnitude for all sources and 5 different subclasses; BL~Lacs and FSRQs (identified according to the presence/size of optical emission lines) and HSP, ISP and LSP sources (classified according to the location of the synchrotron peak in their SEDs). It is evident in the plot of all sources that there are two visible subclasses, both with positive correlations but different ranges in optical and $\gamma$-ray fluxes. These two subclasses are shown to be the FSRQ and BL~Lac sources in the next two plots. So not only do these sources show differences in the strength of their optical emission lines, they also cover different ranges in $\gamma$-ray flux and optical magnitude. The bottom 3 plots in Figure \ref{fig:mag_gam_plot} show that the sources split by the location of their synchrotron peak also cover different ranges in $\gamma$-ray and optical flux. LSP sources are brightest in $\gamma$-ray flux, there is a decrease in maximum $\gamma$-ray flux as the spectral peak moves toward higher frequencies, with the HSP sources in this sample having a much lower range in $\gamma$-ray fluxes compared with LSPs.  

In Table \ref{corrtable} the results of the Spearman Rank analysis of this data are presented and it is demonstrated that (due to differing distances of the sources), the ``whole sample'' approach is not particularly useful. Rather we must consider the properties of each source (which are differently coloured in the figure) individually.  In addition in order to investigate the properties of the individual sources the data were first separated into observing seasons (to avoid false periods of apparently stable behaviour introduced by long periods of non-visibility \citep{itoh2013}). The number of seasons for each sources depends on the availability of the optical data and vary from 1 to 4 seasons.

For each of the season datasets a Spearman Rank Coefficient test was performed to measure the statistical dependence of one flux against the other. A summary of these results are presented, along with those for other correlations, in Table \ref{speartable}. Results for individual source seasons can be found in Table \ref{full_table} in the Appendix. We find that 68\% of source seasons (25/37) show a positive a correlation $\bar{\rho}$~=~0.46 with significant p values (i.e. p$\leq$0.05 or p$\geq$0.95). In addition, 92\% of source seasons (34/37) show positive correlations ($\bar{\rho}$~=~0.36) with p values ranging between 0.000 - 0.988.

% \usepackage{graphicx}
\begin{table*}
\centering
\resizebox{\textwidth}{!}{%
%\begin{tabular}{ll|llllll|llll}
\begin{tabular}{cclcccccclcccc}
\hline
 &  & All P  &  &  &  &  &  & p$\leq$0.05&  &  &  \\ \hline
 & Type & Range p & Range $\rho$ & Mean p & Mean $\rho$ & Quantity& Quantity & Range $\rho$ & Mean $\rho$ & Quantity & Quantity \\ 
&          	&			&	          &		  &    		&-$\rho$,+$\rho$		&		&			&		& -$\rho$,+$\rho$			&		\\ \hline
mag-gam & HSP 	& 2.20x10$^{-16}$ - 0.620 			& -0.0929 - 0.745 	& 0.210	& 0.299 	& 1,6 	& 7 		&   0.502 - 0.745 	& 0.608	& 0,3 	& 3 \\
 		& ISP 	& 4.97x10$^{-11}$ - 0.524			&  -0.067 - 0.718 	& 0.115 	& 0.429 	& 0,16 	& 16 	&   0.287 - 0.718	& 0.567 	& 0,10 	& 10 \\
 		& LSP 	& 0.000 - 0.988		 			&  -0.600 - 0.711 	& 0.141 	& 0.337 	& 2,12 	& 14 	&   -0.600 - 0.711 	& 0.390	& 2,10 	& 12 \\
		& ALL 	& 0.000 - 0.988 					& -0.600	- 0.745 	& 0.142 	& 0.369 	& 3,34 	& 37 	& -0.600 - 0.745 	& 0.487	& 2,23 	& 25 \\ \hline

gam-deg 	 & HSP 	& 1.89x10$^{-13}$ - 0.419 	& -0.121 - 0.633 	& 0.196 	&  0.231		& 1,4 	& 5 		&  0.160  - 0.633 	&  0.397 	& 0,2 	& 2 \\
		 & ISP 	& 6.68x10$^{-4}$ - 0.946 	& -0.560 -  0.411 	& 0.370 	& -0.0382 	& 8,8 	& 16 	& -0.560  -   0.272 	& -0.229 	& 2,1 	& 3 \\
 		& LSP 	& 1.52x10$^{-6}$ - 0.925 	& -0.249 -  0.556 	& 0.340 	&  0.0619 	& 6,5 	& 11 	&  0.360 -   0.556 	&  0.426 	& 0,3 	& 3 \\
 		& ALL 	& 1.89x10$^{-13}$ - 0.946	& -0.560  - 0.633 	& 0.332 	& 0.038 		& 15,17	& 32 	& -0.560  -  0.633 	&  0.173 	& 2,6 	& 8\\ \hline

deg-mag & HSP 	& 1.98x10$^{-5}$ - 0.695 	& 0.0876 - 0.549 			& 0.312 	& 0.268 	& 0,7 	& 7 	& 0.468 - 0.513 	& 0.525 	& 0,2 	& 2 \\
 		& ISP 	& 1.53x10$^{-11}$ - 0.0754 	& -0.485 - 0.395	 		& 0.0212 	& 0.0334 	& 2,2 	& 4 	& -0.460 - 0.403 	& 0.0697 	& 1,2 	& 3 \\
 		& LSP 	& 0.0607 - 0.999 			& 5.47x10${-4}$ - 0.270 	& 0.472 	& 0.154 	& 0,4 	& 4 	& NA 			& NA 	& NA 	& 0 \\
 		& ALL 	& 1.53x10$^{-11}$ - 0.999 	& -0.485  - 0.549 			& 0.277 	& 0.175 	& 2,13 	& 15 & -0.460  - 0.513 	& 0.252 	& 1,4 	& 5 \\
\hline
\end{tabular}
}
\caption{Summary of results from the Spearman Rank correlation test showing the p and $\rho$ values for different subclasses for optical vs $\gamma$-ray data, degree of polarisation vs $\gamma$-ray data and optical flux vs optical degree of polarisation. The full dataset is presented in Table \ref{full_table} in the Appendix.}
\label{speartable}
\end{table*}

A Kolmogorov~Smirnov (KS) test was performed on the HSP \& ISP, ISP \& LSP and HSP \& LSP sources respectively to test whether the distribution of the Spearman Rank Coefficient $\rho$ values for different blazar subclasses suggests that the subclasses originate from the same parent population. The mean of the distributions for each subclass are $\bar{\rho}$~=~0.30, 0.43 and 0.34 for HSP, ISP and LSPs respectively. The p values from the KS test indicate the probability of the HSP and ISP sources being from the same parent population is 56\%, for ISP and LSP sources the probability is 58\% and for the HSP and LSP sources the probability is 84\%. It is not possible to distinguish these probabilities from each other and none has a significant $p$ value to either indicate the subclass results are or are not drawn from the same parent distribution.

%\begin{figure*}
%\centering
%\includegraphics[width=16cm]{zero_lag_R_gam.eps}
%\includegraphics[width=16cm]{updated_hist_zero_lag.eps}
%\caption[]{Histograms showing the distribution of $\rho$ (top) and p (bottom) values from the Spearman Rank Coefficient test for the optical and $\gamma$-ray flux. From left to right: HSPs (green), ISPs (blue) and LSPs (red). The dotted histograms are the distribution of the total sample and the black vertical lines show where the sample mean lies. The mean of the subclasses are shown as a vertical line in their respective colours.}
%\label{fig:lag_compare}
%\end{figure*}

\subsubsection{Optical degree of polarisation and $\gamma$-ray flux correlations}
\begin{figure*}
\centering
\includegraphics[width=18cm]{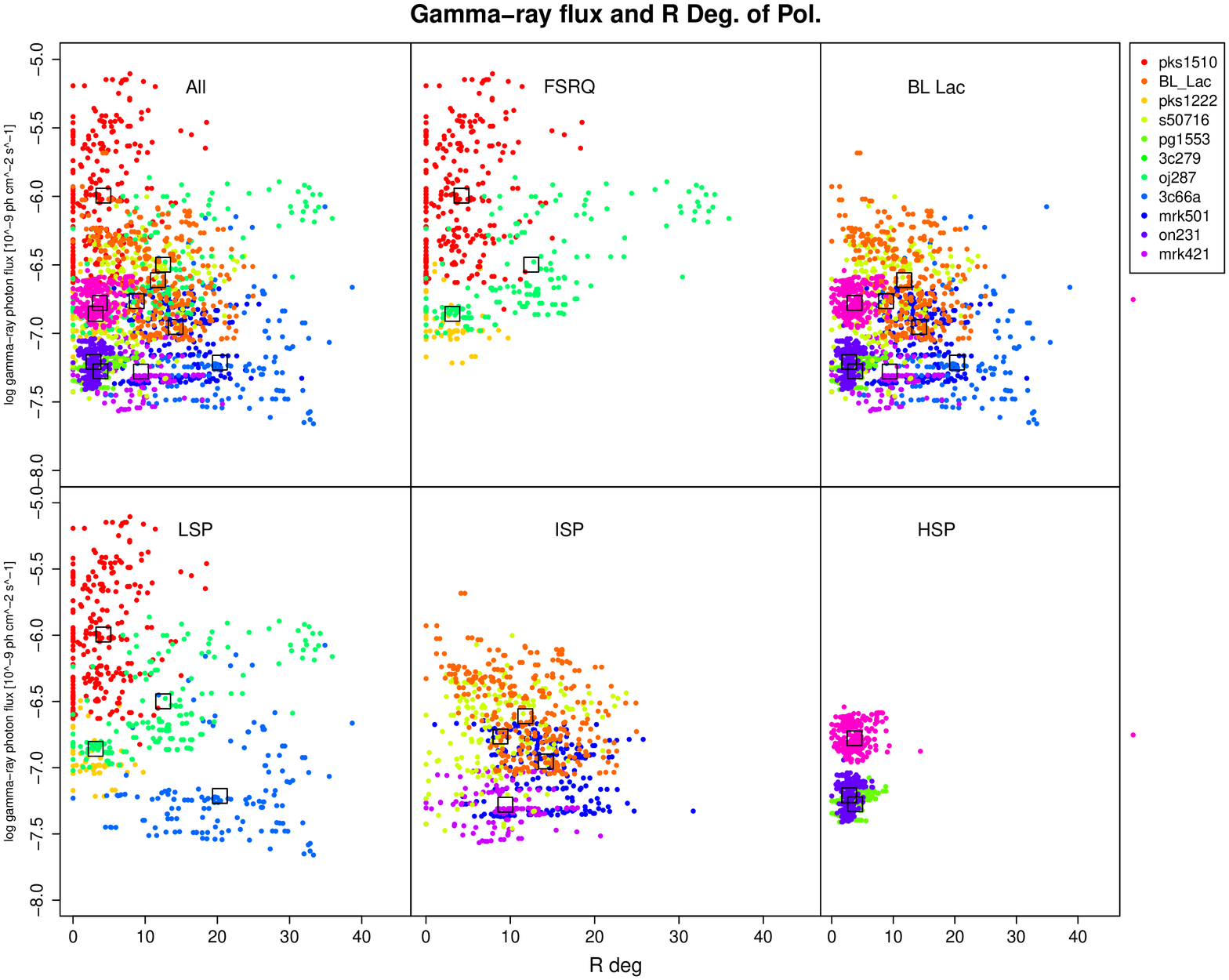}
\caption[]{Fermi $\gamma$-ray flux against optical degree of polarisation for all sources and each blazar subclass, a different colour for each source separately. Black squares show where the mean of the source lies on the plot. Horizontal and vertical lines in the data show periods during which the $\gamma$-ray/optical data (respectively) are constant while the other continues to vary. }
\label{fig:deg_gam_plot}
\end{figure*}

Figure \ref{fig:deg_gam_plot} shows the $\gamma$-ray flux against optical degree of polarisation for all sources (each coloured individually) and the 5 subclasses. The $\gamma$-rays are plotted on a logarithmic scale for visualisation purposes. The horizontal lines are caused by the polarisation varying during a wide $\gamma$-ray flare bin, usually in low $\gamma$-ray states. The vertical lines are caused by the $\gamma$-ray flux varying when the degree of polarisation is very low. For the $\gamma$-ray and degree of polarisation plots it is not possible to distinguish the FSRQ and BL~Lac subclasses from each other. The FSRQs exhibit higher $\gamma$-ray fluxes than the BL~Lacs. The spectral peak subclasses differ in their $\gamma$-ray flux value (as already shown in the previous section), however they also differ in their maximum degree of polarisation value. The LSP sources can exhibit polarisation degrees up to $\sim$40\%, ISPs $\sim$30\%  and the HSP sources have a maximum of $\sim$10\%. The HSP sources also show less variation that the LSP and ISP sources. 

\begin{figure*}
\centering
\includegraphics[width=16cm]{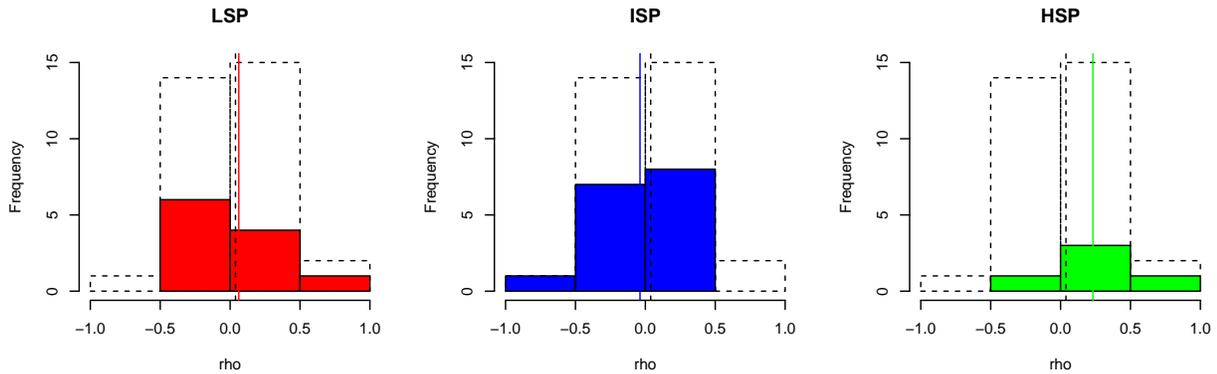}
\caption[]{Histograms showing the distribution of $\rho$ values from the Spearman Rank Coefficient test for the optical degree of polarisation and $\gamma$-ray flux. From left to right: LSPs (red), ISPs (blue) and HSPs (green). The dotted histograms are the distribution of the total sample and the black vertical lines show where the sample mean lies. The mean of the subclasses are shown as a vertical line in their respective colours.}
\label{fig:p_0_hist}
\end{figure*}

In order to investigate the correlations in more detail we again split the data by individual source and by observing season and carried out a Spearman Rank analysis (see summary of results in Table 4). In Figure \ref{fig:p_0_hist} we show the distribution of the $\rho$ coefficients as histograms. The peak of the overall distribution is close to zero (as shown in Table \ref{speartable}. However the peak of the $\rho$ value distributions for the LSP and HSP sources are positive and for ISP sources, negative. All HSP and LSP source seasons show positive correlations with p$\leq$0.05, whereas ISP sources have a slight majority of positive correlations (2/3) with p$\leq$0.05.

We carried out a KS test analysis on the distribution of $\rho$ values. The HSP and ISP distributions have 44\% probability of being from the same parent distribution and the HSP and LSP distribution have a 42\% probability. The probability of the ISP and LSP sources being from the same distribution is 48\%.  None of these p values is significant. 

\subsubsection{Optical flux and degree of polarisation correlations}\label{optopt}
\begin{figure*}
\centering
\includegraphics[width=18cm]{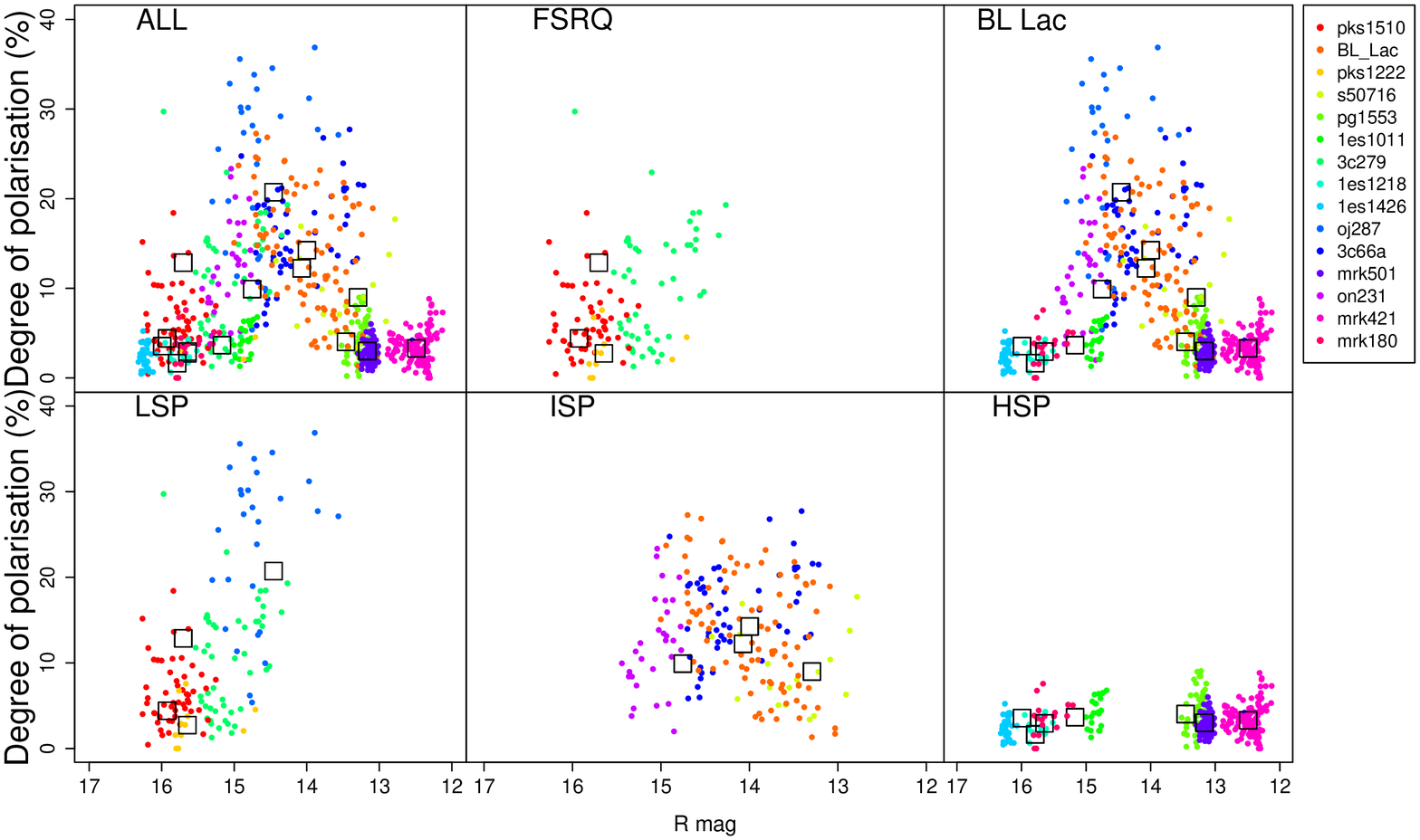}
\caption[]{The optical degree of polarisation against optical magnitude for all 15 sources. For those sources that do not have synchronous points(i.e. Mrk~421 and PKS~1222) we interpolate the nearest lying point from the neighbouring datapoints. Each source is coloured separately and black boxes show where the mean of that source lies on the plot.}
\label{fig:magdeg_gam_plot}
\end{figure*}

Figure \ref{fig:magdeg_gam_plot} shows plots of the degree of polarisation against the optical magnitude separated by object type. Here we plot all 15 sources in our sample (i.e. including those without Fermi data). Those sources that do not have synchronous magnitude and degree of polarisation (Mrk~180 and PKS~1222+216) have their points interpolated from neighbouring data where available. In addition as the data are synchronous they are not split into seasons but compared across the whole available dataset. As already shown in the previous correlation plots, the HSP sources are limited to degree of polarisation values \textless10\%, while the LSP and ISP sources show greater variability, reaching a maximum polarisation of $\sim$40\% and $\sim$30\% respectively. The HSP sources show tighter groupings than the LSP and ISP sources. 

Table \ref{speartable} shows the Spearman Rank Coefficient $\rho$ and probability values for the optical flux and degree of polarisation data. 87\% (13/15) of sources show weak positive correlations between the optical flux and the optical degree of polarisation with $\bar{\rho}$~=~0.18. In addition, 4 sources show weak positive correlations with p$\leq$0.05, however, the HSP sources lack significant correlations (where p$\leq$0.905).

The probability of the HSP, ISP and LSP distributions being from the same parent sample was tested using the KS test. For HSP and ISP sources p~=~0.42, ISP and LSP p~=~0.77 and HSP and LSP p~=~0.66 also. None of these p values is significant and the null hypothesis cannot be rejected.

%\begin{figure*}
%\centering
%\includegraphics[width=16cm]{deg_mag_comp_rhop.eps}
%\caption[]{Histograms of Spearman Rank Coefficient values $\rho$ (top panel) and p (bottom panel) for optical flux against degree of polarisation. For PKS~1222+216 there are only two datapoints which are synchronous in degree of polarisation and optical flux so this source is excluded from the analysis, there is also no synchronous data for Mrk~421 as the photometric and polarimetric data originate from different telescopes. From left to right: HSPs (green), ISPs (blue) and LSPs (red).}
%\label{fig:deg_mag_0_lag}
%\end{figure*}

\subsection{Optical and $\gamma$-ray properties during EVPA rotations}\label{evpaflare}

\begin{figure}
\centering
\includegraphics[angle=270,width=8.5cm]{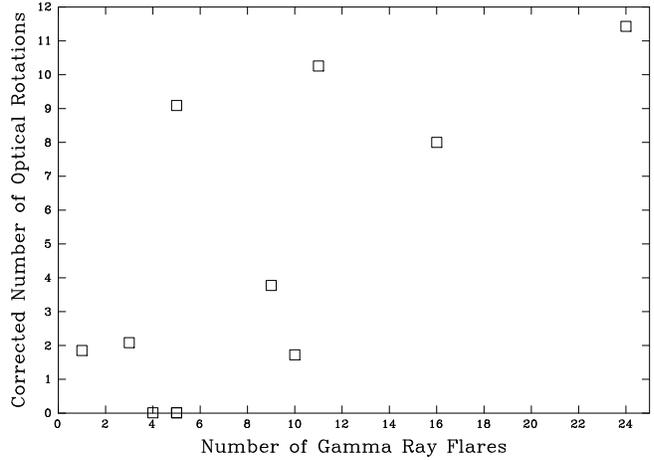}
\caption[]{Number of observed optical rotations (corrected for the observing duty cycle) versus number of observed $\gamma$-ray flares for those sources which have sufficient $\gamma$-ray data (11/15 sources) - note there are two points at x=5, y=0.}
\label{fig:duty}
\end{figure}

We have identified 95 $\gamma$-ray flare events (see Section \ref{description} for the description of a flaring event) in 11 sources.
In the sample, the rate of flaring is between 0.0022 - 0.017 flares per day (0.8 - 6.2 per year). The mean flare rates (and standard deviations) for each subclass are HSP~=~$0.005\pm.001$ , ISP~=~$0.007\pm0.003$ and LSP~=~$0.009\pm0.006$ flares per day, equivalent to HSP~=~2.0$\pm$0.5, ISP~=~2.6$\pm$1.2, LSP~=~3.3$\pm$2.3 flares per year. These results indicate that there are no significant differences between the rate of flaring in the different subclasses.

In order to make a simple assessment if flaring and rotation activity are associated we can compare the number of flares per 
source with the number of rotations.  Since there is missing optical data due to seasonal effects,
we correct the number of rotations of a given source by dividing by its optical duty cycle (defined as the fraction of time when optical coverage overlapped with the Fermi data).  The results of this analysis are presented in  Figure \ref{fig:duty}.  A significant correlation ($\rho=0.59, p=0.05)$ is apparent.  It therefore appears that there is at least some link between a propensity for $\gamma$-ray flaring and that for optical polarisation rotations.

Due to the visibility of the sources, 67 of the $\gamma$-ray flaring periods occur when we lack coincident optical data or there are no data between the data point and the nearest flare. Of the remaining 28 $\gamma$-ray flares that have optical photometry and polarimetry in coincident periods with the $\gamma$-ray data there are 17 that occur during rotation of the EVPA (see Table 2, Column 10).  However we note that this statistic is dominated by one source (PKS~1510-089) which has the highest mean flare rate and multiple flares within a single long EVPA rotation.  In addition we can associate 11 flares that occur outside an EVPA rotation with the closest in time EVPA rotation (i.e. the nearest lying rotation to a flare where there are no missing data in between). There are 5 flares that occurred \textless84 days after the rotation and 6 flares that occurred \textless116 days before the rotation (see Table 2, Column 9). Even though we do not analyse flares that occur during periods when we lack optical data, we must be cautious: the average observing season is $\sim$180 days which means that it may be possible that flares could be associated with closer lying rotations that occur when we are unable to observe them.

\begin{figure*}
\centering
\includegraphics[width=15cm]{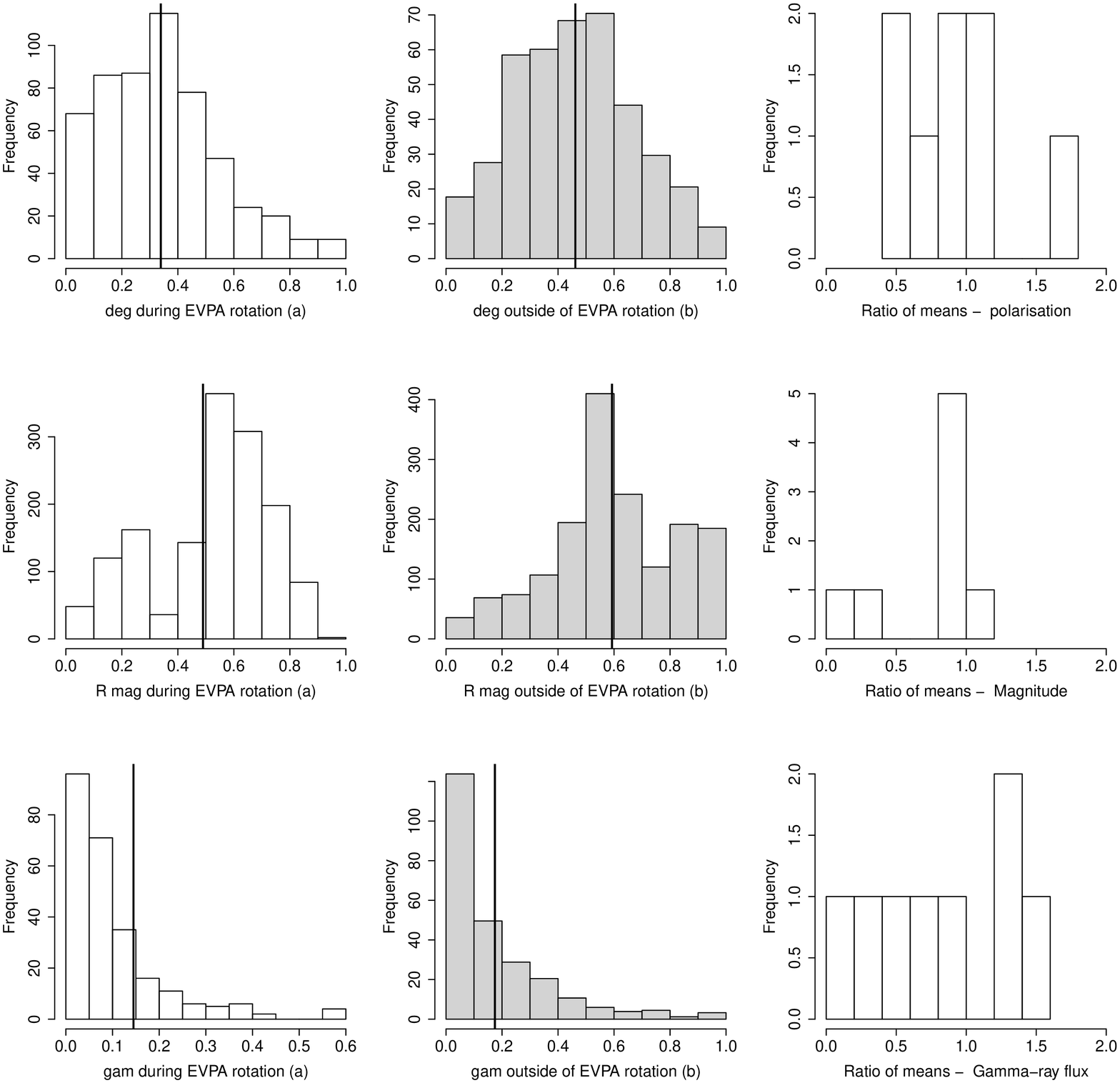}
\caption[]{The degree of polarisation (top), optical magnitude (middle) and $\gamma$-ray flux (bottom) displayed as a fraction of the normalised range for a) during EVPA rotations (white) and b) outside of EVPA rotations (grey) and c) as a ratio of the mean of each property during a rotation over the mean of each property outside of a rotation for each individual source (see Section \ref{evpaflare} for more details). The black vertical lines in the first two columns show the mean of the histograms.}
\label{fig:dop_during_rot}
\end{figure*}

In order to investigate the $\gamma$-ray and optical properties during and outside of rotations we separated the data for each source into two periods: during (a) rotation and (b) non-rotation. The first two histograms in the top panel of Figure \ref{fig:dop_during_rot} show the degree of polarisation for all sources during those periods.  The data are presented as a percentage of the full range of the degree of polarisation for a particular source and the whole histogram has been divided by the ratio of the number of points in the larger dataset (outside of EVPA rotations) over the number of points in the smaller dataset (during EVPA rotations), this removes rare events from the analysis and takes into account any selection effects. After this normalising we find that the distributions do not change and each bin still has $\geq$1 occurrence.

The top panel shows the distribution of the degree of polarisation during a rotation is generally shifted toward lower values and the high polarisation tail is suppressed. Outside of the rotations the data appears to have a more Gaussian distribution. The mean of the distributions of degree of polarisation ($\bar{DoP}$) during a rotation is 0.34 and outside of a rotation $\bar{DoP}$~=~0.46.  On average the degree of polarisation is therefore 26\% lower during a rotation.  A KS test was performed on the data to establish the probability that the degree of polarisation during rotation and non-rotation events are from the same parent distribution. There is a very low probability (p\textless0.01\%) that the degree of polarisation during rotations comes from the same distribution as the degree of polarisation during non-rotations. The null hypothesis is rejected and we can state that the distribution of the degree of polarisation is  different during rotations to outside of rotations.

In the middle panel of Figure \ref{fig:dop_during_rot} the first two histograms show the R magnitude, during EVPA rotations and outside EVPA rotations, as a fraction of the normalised range of the total magnitude. Similar to the degree of polarisation, in very relatively few points does the magnitude reach \textgreater90\% of the total flux during a rotation whereas outside of rotations the are $\sim$300 points that have polarisation values \textgreater90\% of the total flux. The mean of the distribution during a rotation is $\bar{R}$ = 52\% and outside of rotation periods the magnitude is $\bar{R}$ = 59\%. On average the degree of polarisation is therefore 17\% lower during a rotation. The results from the KS test show, as for the degree of polarisation, the two distributions of magnitude during and outside of rotation periods have a very low (p\textless0.01\%) probability of being from the same distribution. The null hypothesis is again accepted; during a rotation the optical flux is lower compared to outside of a rotation and their distributions are not from the same initial sample.

%\begin{figure*}
%\centering
%\includegraphics[width=20cm]{mag_during_rot_comp.eps}
%\caption[]{The median optical flux displayed as a value between 0 and 1 of the normalised range for a) during rotations and b) during no rotations.}
%\label{fig:mag_during_rot}
%\end{figure*}

The first two histograms in the bottom panel of Figure \ref{fig:dop_during_rot} show the relative strength of the $\gamma$-ray flux during- and outside of- EVPA rotations. During the rotations the $\gamma$-ray flux never rises above 59\% of the total $\gamma$-ray flux. Outside of the rotations the $\gamma$-ray flux has a longer high-$\gamma$ flux tail, with the maximum brightness occurring outside of a rotation event. The KS test results show that the likelihood of the rotation and non-rotation $\gamma$-ray flux to be from the same parent population is 24\%. This means the null hypothesis, that the samples are from the same distribution, cannot be formally rejected. However we note that during rotations, the mean of the $\gamma$-ray flux distribution ($\bar{\gamma}$ = 10\%) is 42\% lower compared to that outside of a rotation ($\bar{\gamma}$ = 17\%). 

The third column in Figure 8 shows, for each individual source, the mean ratio of degree of polarisation (top), R magnitude (middle) and γ-ray flux (bottom) during and outside of a rotation. For the degree of polarisation, 5/8 sources have lower values (i.e. ratios \textless1) during rotations. For the R magnitude there are 7/8 sources that have lower values during a rotation. For the $\gamma$-rays there are 5/8 sources which are less bright in $\gamma$-rays during a rotation.

%\begin{figure*}
%\centering
%\includegraphics[width=20cm]{gam_during_rot_comp.eps}
%\caption[]{The median $\gamma$-ray flux displayed as a value between 0 and 1 of the normalised range for a) during rotations and b) during no rotations.}
%\label{fig:gam_during_rot}
%\end{figure*}

\section{Conclusions}\label{conc}
There are important caveats to consider before making concluding remarks about the results presented in this paper. Firstly, the RINGO2 blazar monitoring survey was designed to follow-up sources detected by MAGIC, the original sample size has increased, but the essence of the sample is that the sources are all $\gamma$-ray bright and have exhibited some kind of flaring activity (hence the reason they are added to the sample). Thus, due to this selection bias, the presented sample averaged results (Section 4) cannot be generalised to the larger blazar population, however the correlations for individual sources (Section 3) are robust.

In Section 3 we presented a detailed discussion of the behaviour of the individual sources in our sample.  Comparing source to source the orientation of the rotation (i.e. whether it is upward or downward) does not afford any information as the rotation direction is presumably subject to the arbitrary sense of the magnetic field and its properties which will vary from blazar to blazar. However, in four sources 3C279, PKS~1510-089, PG~1553+113 and S5~0716+714 we observe upward then downward rotations and in the case of S5~0716+714 we see the EVPA rotate upward, downward and then upward again (see light curves for these sources in the Appendix). We also have cases in four sources PKS~1510-089, S5~0716+714, PG~1553+113 and Mrk~421 in which there is a $\gamma$-ray flare, during or temporally close, associated with a rotation. Such behaviour of the EVPA is potentially important in studying the magnetic field and/or the orientation of the jet/emission blob within the jet with respect to the observer. Monte Carlo analysis by \cite{blinov2015} suggests that a single EVPA rotation event can be caused by a random walk of the EVPA, but it was unlikely that {\it all} rotations are due to random walk.

In Section 4 we carried out a statistical study of the general properties of these sources without considering their individual behaviour.  We found the following principal results:

\begin{enumerate}
\item The maximum observed degree of optical polarisation for the LSP sources was $\sim40$\%.  For ISP sources it was $\sim$30\% and for HSP sources $\sim10$\%. It is natural to attribute the low maximum polarisation degree in HSP sources to their optical light being dominated by non-synchrotron emission which could originate from the accretion disk or emitting regions outside of the jet.  This explanation also accords with the low optical variability in these sources, however it could also be a signature of low-ordered magnetic fields in the jet. It must also be noted that these results cannot be applied to the larger blazar population.
\item On average the optical degree of polarisation and $\gamma$-ray flux are not strongly correlated. ISP and LSP sources show no strong preference for either positive or negative correlations. HSP sources show a stronger (yet still weak) positive correlation. %We note that the optical flux wavelength range for the instruments in this study is closest to the wavelengths at which ISP sources peak, this may provide a natural explanation for the positive correlation in ISPs if their optical light is dominated by synchrotron emission.  
\item In 92\% (34/37) of source seasons we found a positive correlation ($\bar{\rho}$~=~0.37) between optical and $\gamma$-ray flux. In over half of the seasons (25/37 = 68\%) the probability of correlation is significant (i.e. p$\leq$0.05). Similar findings have also been reported by \cite{hovatta2014}) and \cite{cohen2014}. Such behaviour may provide evidence of a close physical association between the optical and $\gamma$-ray emitting regions in blazars. We find no significant evidence to determine if the HSP, ISP and LSP distributions of the correlation coefficient were similar or not. This suggests, on average, a common mechanism connects $\gamma$-ray flaring and optical polarisation in these different blazar subclasses.
\item There is a weak positive ($\bar{\rho} = 0.18$) correlation between optical flux and degree of polarisation in 13/15 source seasons.  In 5/15 cases the probability of correlation ($\bar{\rho}$~=~0.25) is significant (i.e. p$\leq$0.05).
\item All blazar subclasses show $\gamma$-ray flaring and EVPA rotations. There is a significant correlation ($\rho$~=~0.59, p~=~0.05) between the number of flares and the number of EVPA rotations in a given object. We do not, however, find any systematic difference by class in $\gamma$-ray flaring rate or number of EVPA rotations. 
%\item We find no significant difference in $\gamma$-ray flare rates between the different blazar subclasses.  We also see EVPA rotations in all subclasses.
%\item A significant correlation ($\rho=0.59, p=0.05)$ is apparent between the number of flares from an object an the number of EVPA rotations.  It therefore appears that there is at least some link between a propensity for $\gamma$-ray flaring and that for optical polarisation rotations.  
\item $\gamma$-ray flaring episodes can occur during and outside of rotation events.  The distribution of lead and lag values between flares and rotations show that there is no preference for either behaviour. The association of the $\gamma$-ray flare and the EVPA rotation could provide evidence for the cause of the rotation and the flare originating from the same shock region, whereby the shock provides electrons for up-scattering photons via Inverse-Compton processes and the tangled magnetic field providing the structure for the EVPA rotation. However, optical and gamma-ray flaring is not always synchronous with an observed rotation, suggesting that other mechanisms are involved in some instances.
\item The mean degree of polarisation as a percentage of the total range of polarisation is 26\% lower during periods of rotation compared to periods of non-rotation, \cite{blinov2016} also report a decrease in polarisation during rotations. The mean optical flux is 17\% lower during a rotation compared with outside rotations and the mean $\gamma$-ray flux is 41\% lower during a rotation compared with outside a rotation. The lower degree of polarisation during a rotation can be interpreted as a difference in the degree of ordering of the magnetic field during a rotation compared with non-rotation. Alternatively it could be evidence for their association with emission features or shocks travelling along helical magnetic field lines \citep{marscher2008,zhang2015}.
\end{enumerate}

\section{Acknowledgments}
We would like to thank the reviewer for their careful reading of the paper and constructive comments. H. Jermak is supported by the Science and Technology Facilities Council (STFC) funding and a Royal Astronomical Society grant. T. Hovatta was supported by the Academy of Finland project number 267324. C. Mundell acknowledges support from the Royal Society, the Wolfson Foundation and the STFC. UBA is partially funded by a CNPq Research Productivity grant number 309606/2013-6 from the Ministry of Science, Technology and Innovation of Brazil. We thank Asaf Pe'er for fruitful discussions during his visit to the ARI, also theoretical discussions with Shiho Kobayashi and Drejc Kopac and statistical discussions with Chris Collins.  The authors thank Benoit Lott for providing the adaptive binning light curve analysis code for the Fermi Gamma-Ray Telescope data. The Liverpool Telescope is operated on the island of La Palma by Liverpool John Moores University in the Spanish Observatorio del Roque de los Muchachos of the Instituto de Astrofisica de Canarias with financial support from the UK STFC.

\clearpage
\bibliographystyle{mn2e}

\bibliography{Papers/Paper}

\section{Appendix}
We present here the fifteen light~curves covering the RINGO2 period of monitoring. Along with the light~curves we discuss the historical behaviour of the sources and how this is relates or differs from the RINGO2 observations. The four windows show (from top to bottom) Fermi $\gamma$-ray data, optical EVPA, optical degree of polarisation and optical flux density. In the Fermi window the areas of rotations are shown (pink for upwards rotation, green for downwards rotation), along with the areas that lack corresponding optical polarisation data (grey) and the Fermi flares (blue) identified using the automated code. For the core sample of 8 sources along with Mrk~421 a large quantity of polarimetric observations are available in the literature. We have reviewed observations from catalogues as well as papers dedicated to single sources and compare our data with the historical behaviour. We describe the $\gamma$-ray emission as High Energy (HE: E\textgreater100 MeV) or Very High Energy (VHE: E\textgreater100 GeV) regimes. 

\begin{landscape}
\begin{figure}
\centering
\includegraphics[width=22cm]{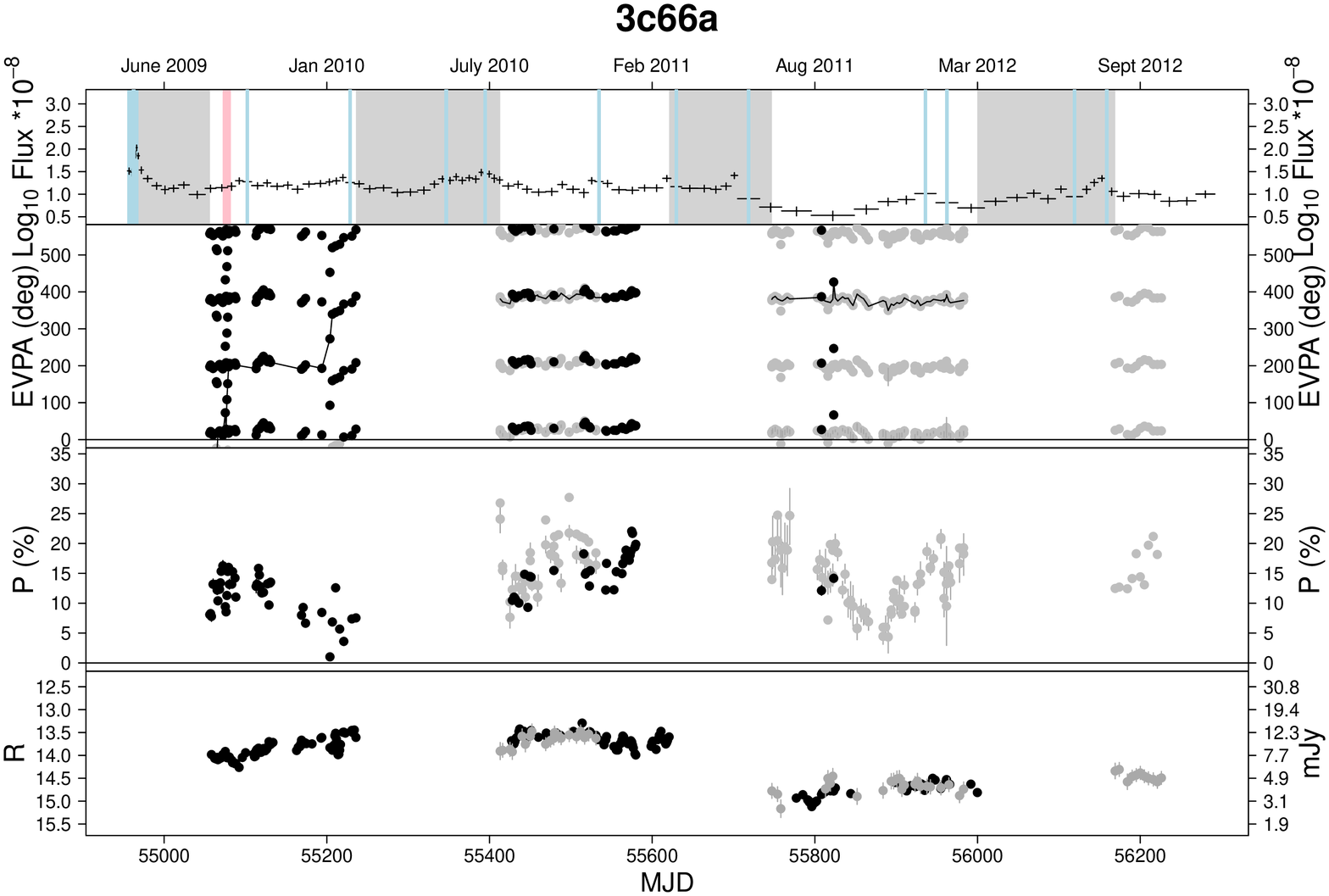}
\caption[3c66a_lcurves]{All $\gamma$-ray and optical data for 3C~66A. Top panel shows the Fermi $\gamma$-ray light curve. The errors on the x axis represent the bins used for the Fermi data. Grey vertical sections show periods where no synchronous optical data available, pink vertical sections highlight regions where optical polarisation angle rotations occur in the upwards direction, light green sections show downward rotations. Flaring episodes are identified by vertical blue lines (see Section \ref{GRF} for definition of a flare). The second panel shows the optical polarisation angle or electric vector position angle (EVPA), the grey points are RINGO2 data and the black points KVA-DIPOL data. The black line traces the temporally closest EVPA points, showing the most likely behaviour of the EVPA. The third panel shows the optical degree of polarisation, and the fourth panel the optical magnitude, all point colours are the same as those for panel 2.}
\label{fig:3c66a}
%\end{sidewaysfigure}
\end{figure}
\end{landscape}

\begin{landscape}
\begin{figure}
\centering
\includegraphics[width=22cm]{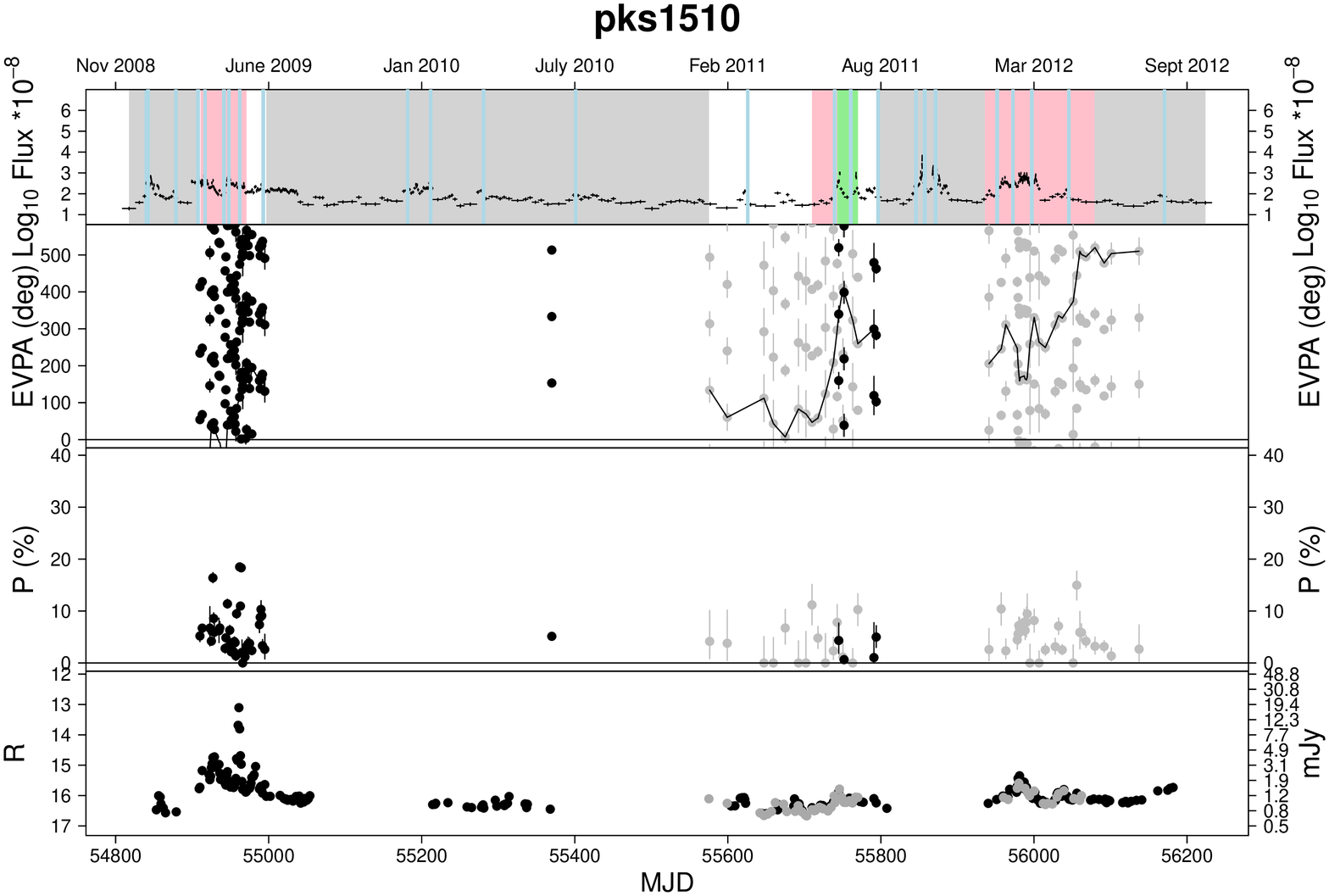}
\caption[pks1510_lcurves]{All $\gamma$-ray and optical data for PKS~1510-089. Top panel shows the Fermi $\gamma$-ray light curve. The errors on the x axis represent the bins used for the Fermi data. Grey vertical sections show periods where no synchronous optical data available, pink vertical sections highlight regions where optical polarisation angle rotations occur in the upwards direction, light green sections show downward rotations. Flaring episodes are identified by vertical blue lines (see Section \ref{GRF} for definition of a flare). The second panel shows the optical polarisation angle or electric vector position angle (EVPA), the grey points are RINGO2 data and the black points KVA-DIPOL data. The black line traces the temporally closest EVPA points, showing the most likely behaviour of the EVPA. The third panel shows the optical degree of polarisation, and the fourth panel the optical magnitude, all point colours are the same as those for panel 2.}
\label{fig:pks1510}
\end{figure}
\end{landscape}

\begin{landscape}
\begin{figure}
\centering
\includegraphics[width=22cm]{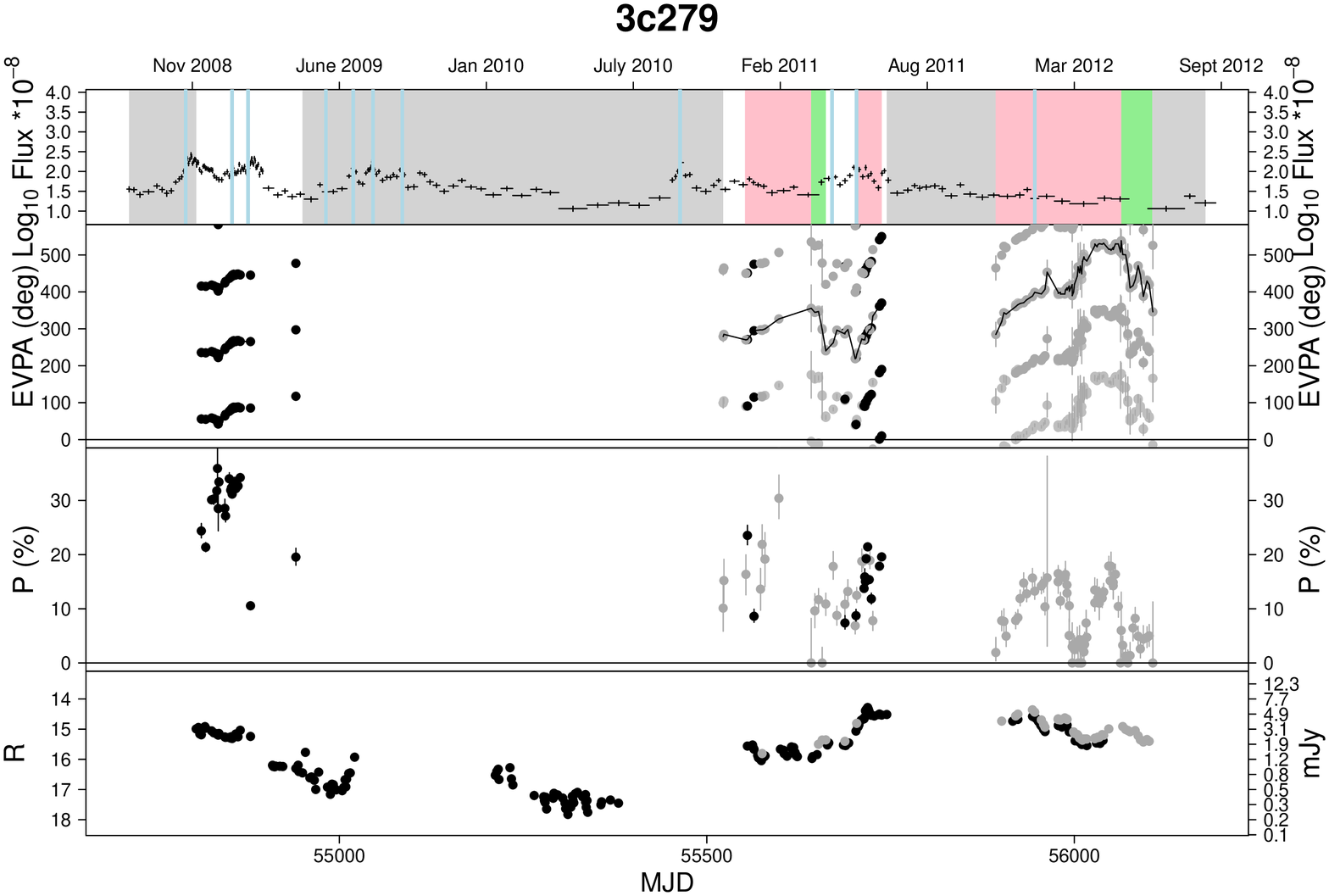}
\caption[3c279_lcurves]{All $\gamma$-ray and optical data for 3C~279. Top panel shows the Fermi $\gamma$-ray light curve. The errors on the x axis represent the bins used for the Fermi data. Grey vertical sections show periods where no synchronous optical data available, pink vertical sections highlight regions where optical polarisation angle rotations occur in the upwards direction, light green sections show downward rotations. Flaring episodes are identified by vertical blue lines (see Section \ref{GRF} for definition of a flare). The second panel shows the optical polarisation angle or electric vector position angle (EVPA), the grey points are RINGO2 data and the black points KVA-DIPOL data. The black line traces the temporally closest EVPA points, showing the most likely behaviour of the EVPA. The third panel shows the optical degree of polarisation, and the fourth panel the optical magnitude, all point colours are the same as those for panel 2.}
\label{fig:3c279}
\end{figure}
\end{landscape}

\begin{landscape}
\begin{figure}
\centering
\includegraphics[width=22cm]{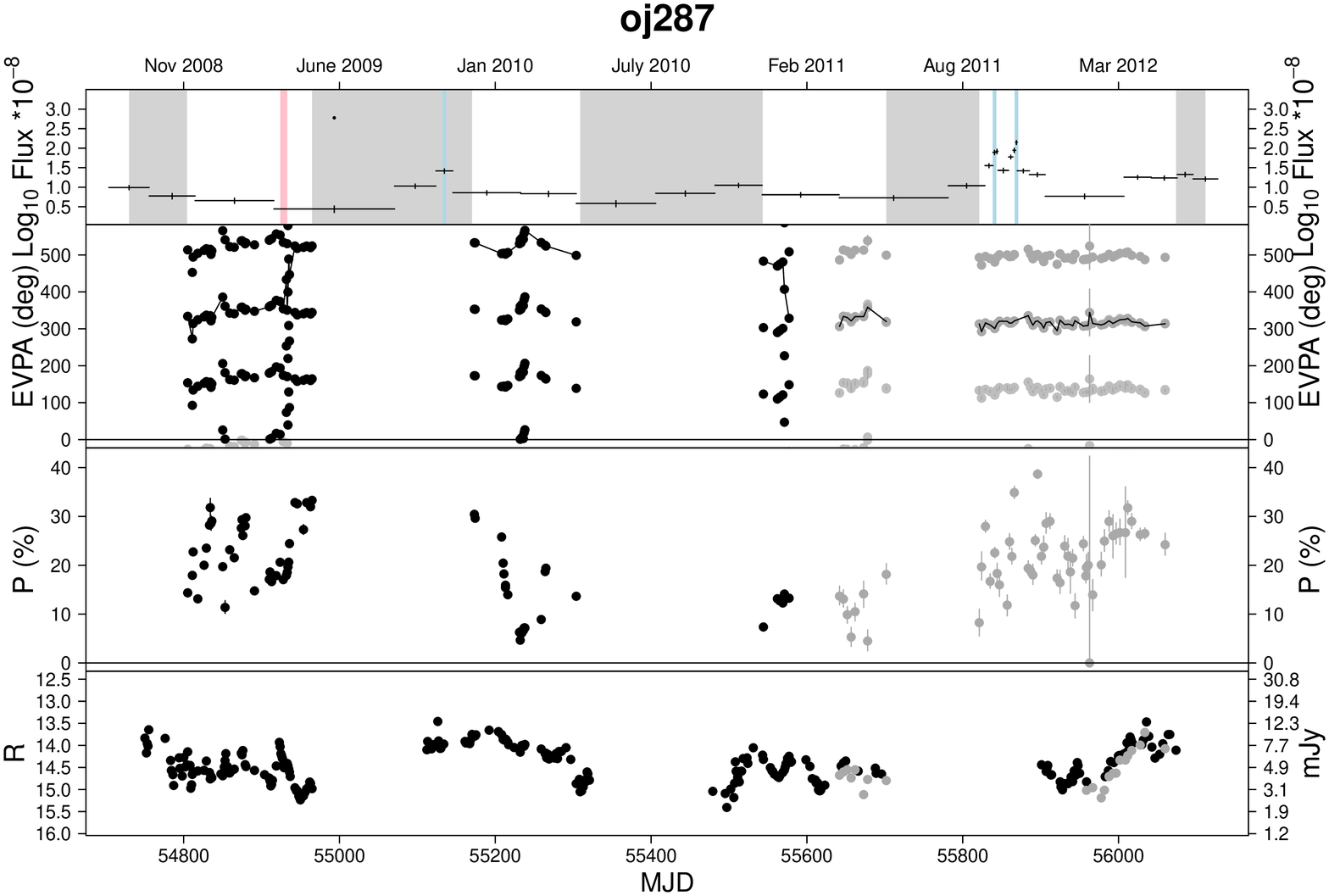}
\caption[oj287_lcurves]{All $\gamma$-ray and optical data for OJ287. Top panel shows the Fermi $\gamma$-ray light curve. The errors on the x axis represent the bins used for the Fermi data. Grey vertical sections show periods where no synchronous optical data available, pink vertical sections highlight regions where optical polarisation angle rotations occur in the upwards direction, light green sections show downward rotations. Flaring episodes are identified by vertical blue lines (see Section \ref{GRF} for definition of a flare). The second panel shows the optical polarisation angle or electric vector position angle (EVPA), the grey points are RINGO2 data and the black points KVA-DIPOL data. The black line traces the temporally closest EVPA points, showing the most likely behaviour of the EVPA. The third panel shows the optical degree of polarisation, and the fourth panel the optical magnitude, all point colours are the same as those for panel 2.}
\label{fig:oj287}
\end{figure}
\end{landscape}

\begin{landscape}
\begin{figure}
\centering
\includegraphics[width=22cm]{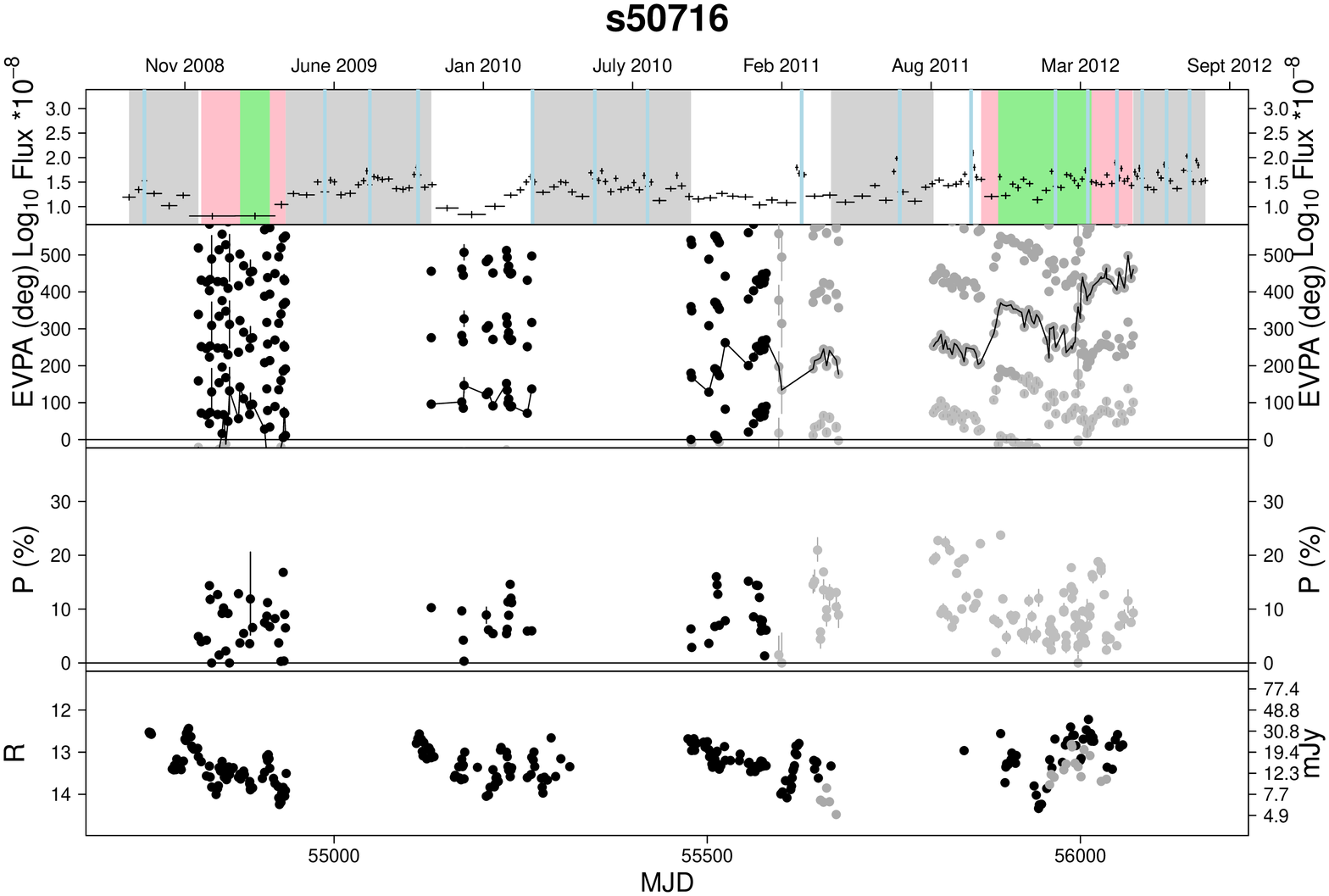}
\caption[s50716_lcurves]{All $\gamma$-ray and optical data for S5~0716. Top panel shows the Fermi $\gamma$-ray light curve. The errors on the x axis represent the bins used for the Fermi data. Grey vertical sections show periods where no synchronous optical data available, pink vertical sections highlight regions where optical polarisation angle rotations occur in the upwards direction, light green sections show downward rotations. Flaring episodes are identified by vertical blue lines (see Section \ref{GRF} for definition of a flare). The second panel shows the optical polarisation angle or electric vector position angle (EVPA), the grey points are RINGO2 data and the black points KVA-DIPOL data. The black line traces the temporally closest EVPA points, showing the most likely behaviour of the EVPA. The third panel shows the optical degree of polarisation, and the fourth panel the optical magnitude, all point colours are the same as those for panel 2.}
\label{fig:s50716}
\end{figure}
\end{landscape}

\begin{landscape}
\begin{figure}
\centering
\includegraphics[width=22cm]{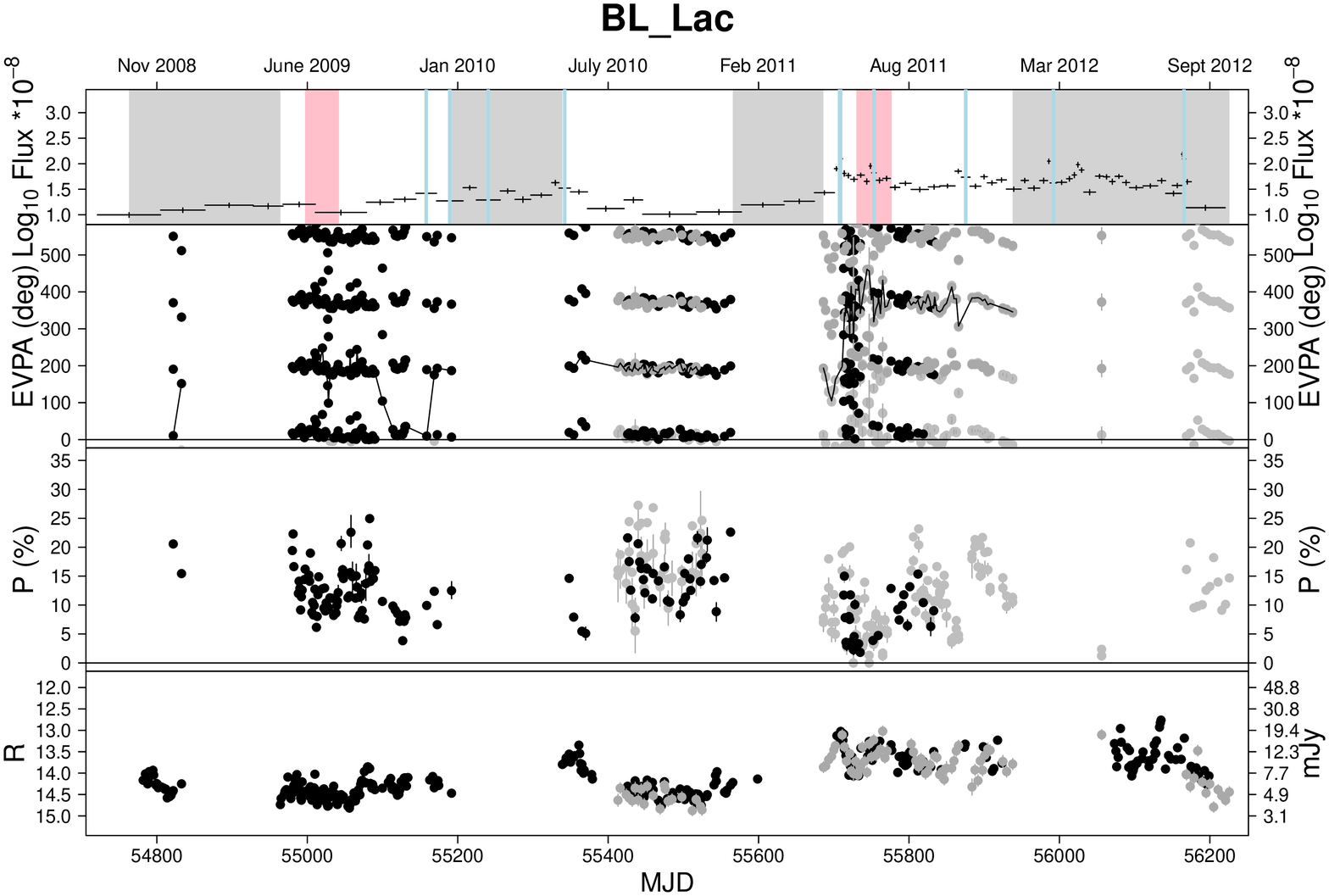}
\caption[bllac_lcurves]{All $\gamma$-ray and optical data for BL~Lac. Top panel shows the Fermi $\gamma$-ray light curve. The errors on the x axis represent the bins used for the Fermi data. Grey vertical sections show periods where no synchronous optical data are available, pink vertical sections highlight regions where optical polarisation angle rotations occur in the upwards direction, light green sections show downward rotations. Flaring episodes are identified by vertical blue lines (see Section \ref{GRF} for definition of a flare). The second panel shows the optical polarisation angle or electric vector position angle (EVPA), the grey points are RINGO2 data and the black points KVA-DIPOL data. The black line traces the temporally closest EVPA points, showing the most likely behaviour of the EVPA. The third panel shows the optical degree of polarisation, and the fourth panel the optical magnitude, all point colours are the same as those for panel 2.}
\label{fig:bllac}
\end{figure}
\end{landscape}

\begin{landscape}
\begin{figure}
\centering
\includegraphics[width=22cm]{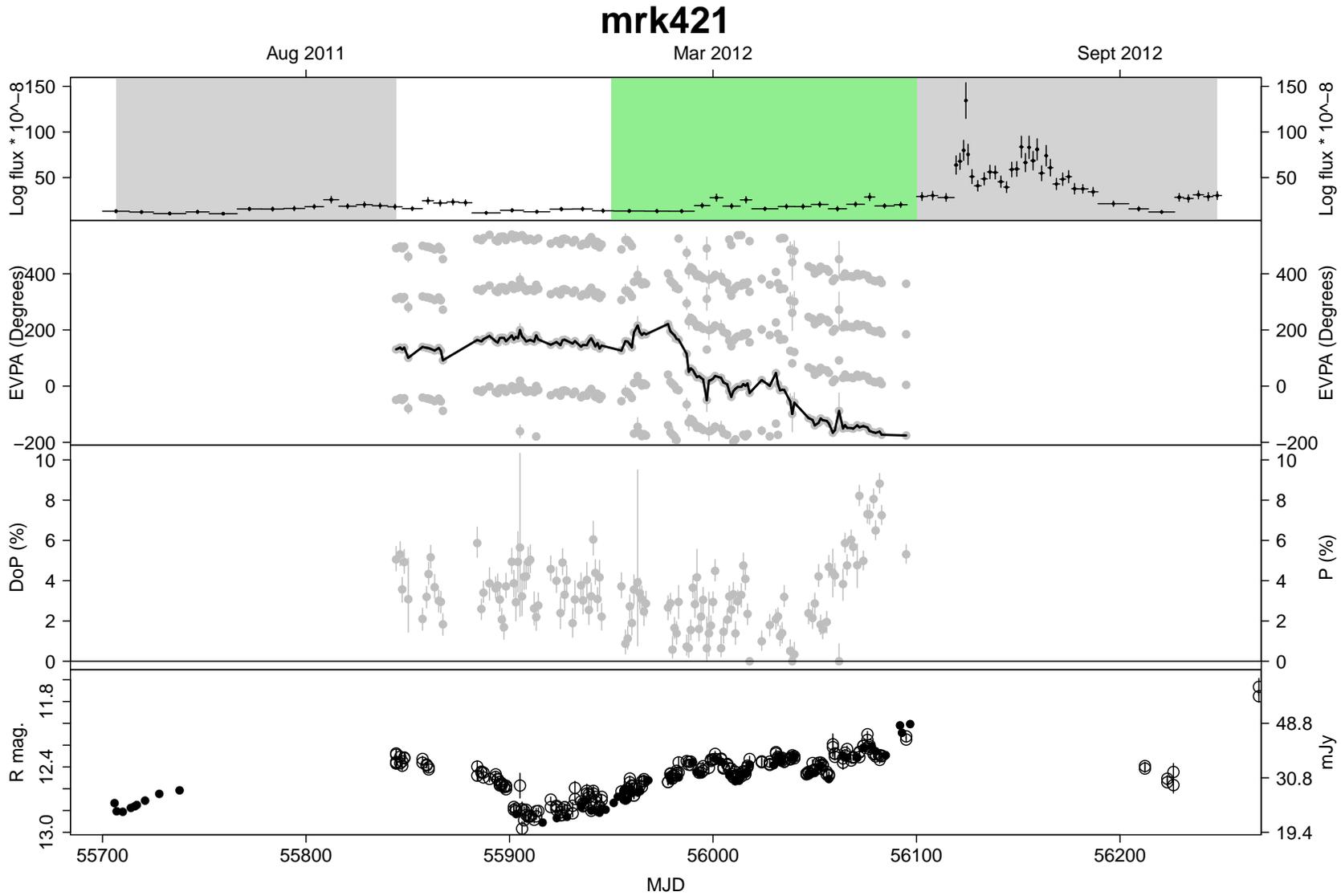}
\caption[mrk421_lcurves]{All $\gamma$-ray and optical data for 3Mrk~421 Top panel shows the Fermi $\gamma$-ray light curve. The errors on the x axis represent the bins used for the Fermi data. Grey vertical sections show periods where no synchronous optical data available, the green vertical section highlights the region where the optical polarisation angle rotates in the downwards direction. Flaring episodes are identified by vertical blue lines (see Section \ref{GRF} for definition of a flare). The second panel shows the optical polarisation angle or electric vector position angle (EVPA), the grey points are RINGO2 data and no KVA-DIPOL data are available. The black line traces the temporally closest EVPA points, showing the most likely behaviour of the EVPA. The third panel shows the optical degree of polarisation (again no KVA-DIPOL data are available), and the fourth panel shows the optical magnitude; photometric calibration of the RINGO2 data was not possible due to the lack of suitable secondary stars in the frame, we instead present SkyCamZ data (open circles) to complement the KVA-DIPOL data.}
\label{fig:mrk421}
\end{figure}
\end{landscape}

\begin{landscape}
\begin{figure}
\centering
\includegraphics[width=22cm]{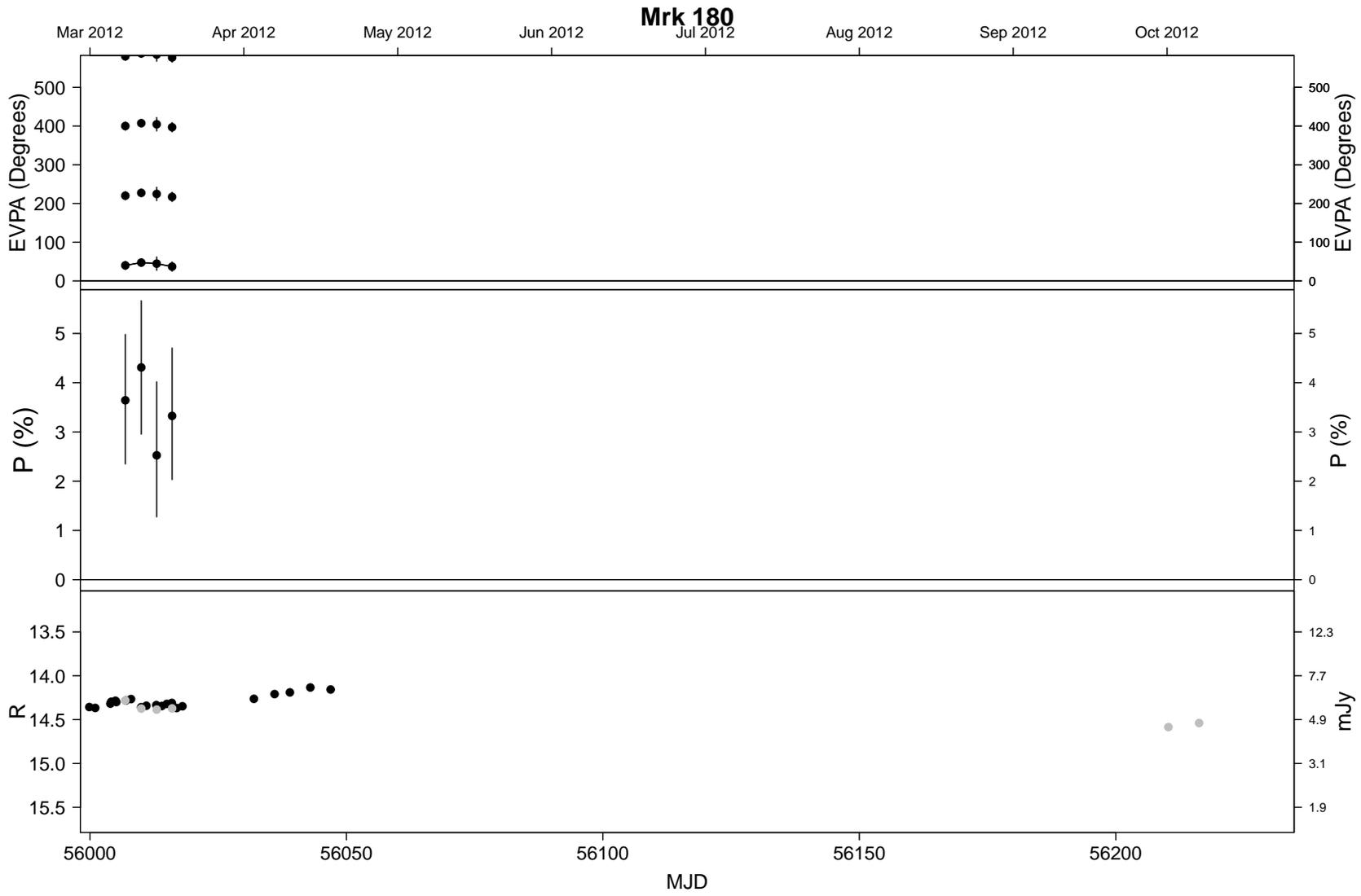}
\caption[mrk180_lcurves]{All optical data for Mrk~180 (the source is too faint in Fermi data). Top panel shows the optical polarisation angle or electric vector position angle (EVPA), the grey points are RINGO2 data and there are no polarisation data points from KVA-DIPOL. The second panel shows the optical degree of polarisation, and the third panel the optical magnitude, all point colours are the same as those for Figure \ref{fig:mrk421}.}
\label{fig:mrk180}
\end{figure}
\end{landscape}

\begin{landscape}
\begin{figure}
\centering
\includegraphics[width=22cm]{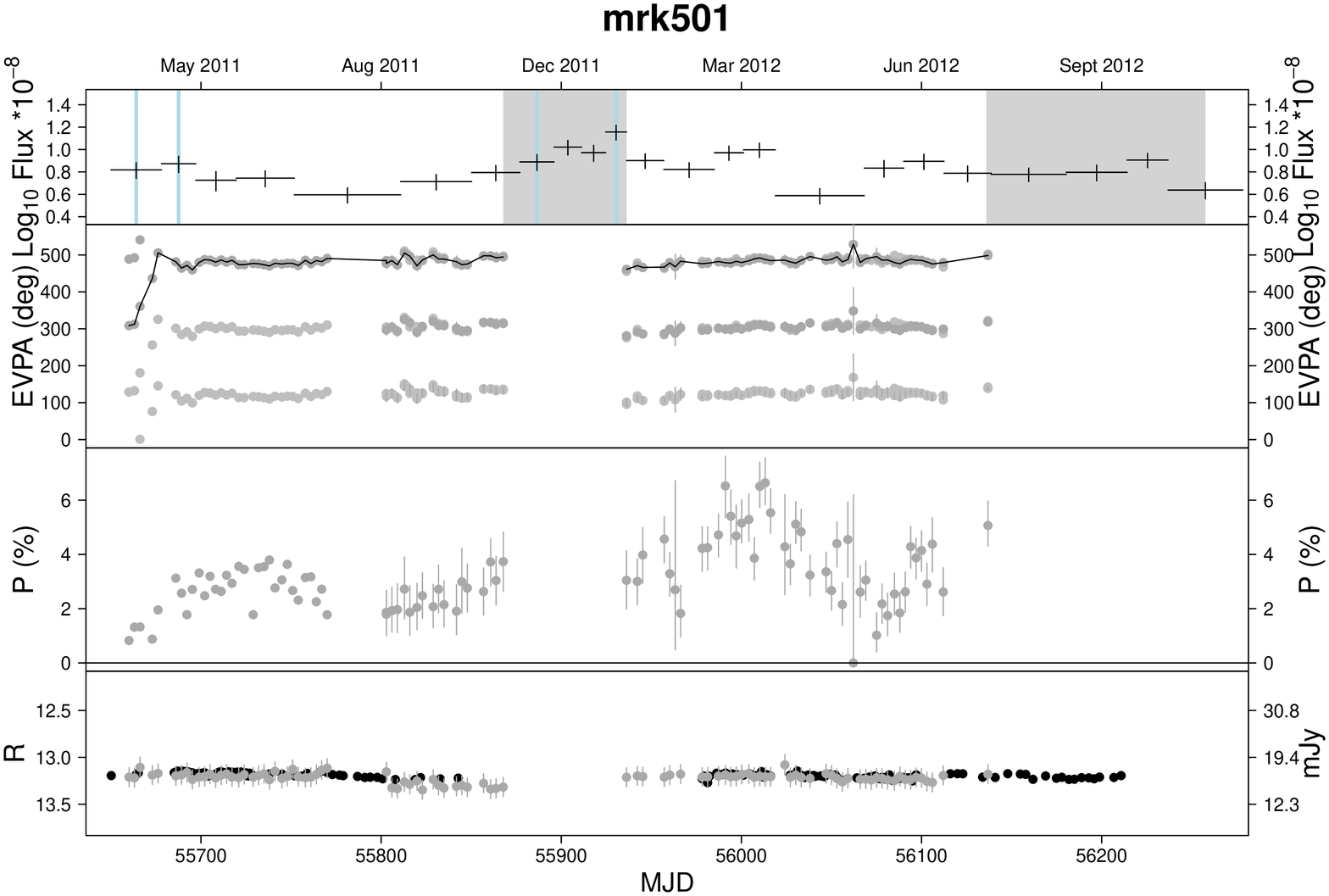}
\caption[mrk501_lcurves]{All $\gamma$-ray and optical data for Mrk~501. Top panel shows the Fermi $\gamma$-ray light curve. The errors on the x axis represent the bins used for the Fermi data. Grey vertical sections show periods where no synchronous optical data are available. Flaring episodes are identified by vertical blue lines (see Section \ref{GRF} for definition of a flare). The second panel shows the optical polarisation angle or electric vector position angle (EVPA), the grey points are RINGO2 data, there are no KVA-DIPOL data for this source. The black line traces the temporally closest EVPA points, showing the most likely behaviour of the EVPA. There are no EVPA rotations (i.e. \textgreater90$^\circ$). The third panel shows the optical degree of polarisation (grey points are RINGO2 and no KVA-DIPOL data available). The fourth panel shows the optical magnitude (black points KVA-DIPOL, grey points RINGO2).}
\label{fig:mrk501}
\end{figure}
\end{landscape}

\begin{landscape}
\begin{figure}
\centering
\includegraphics[width=22cm]{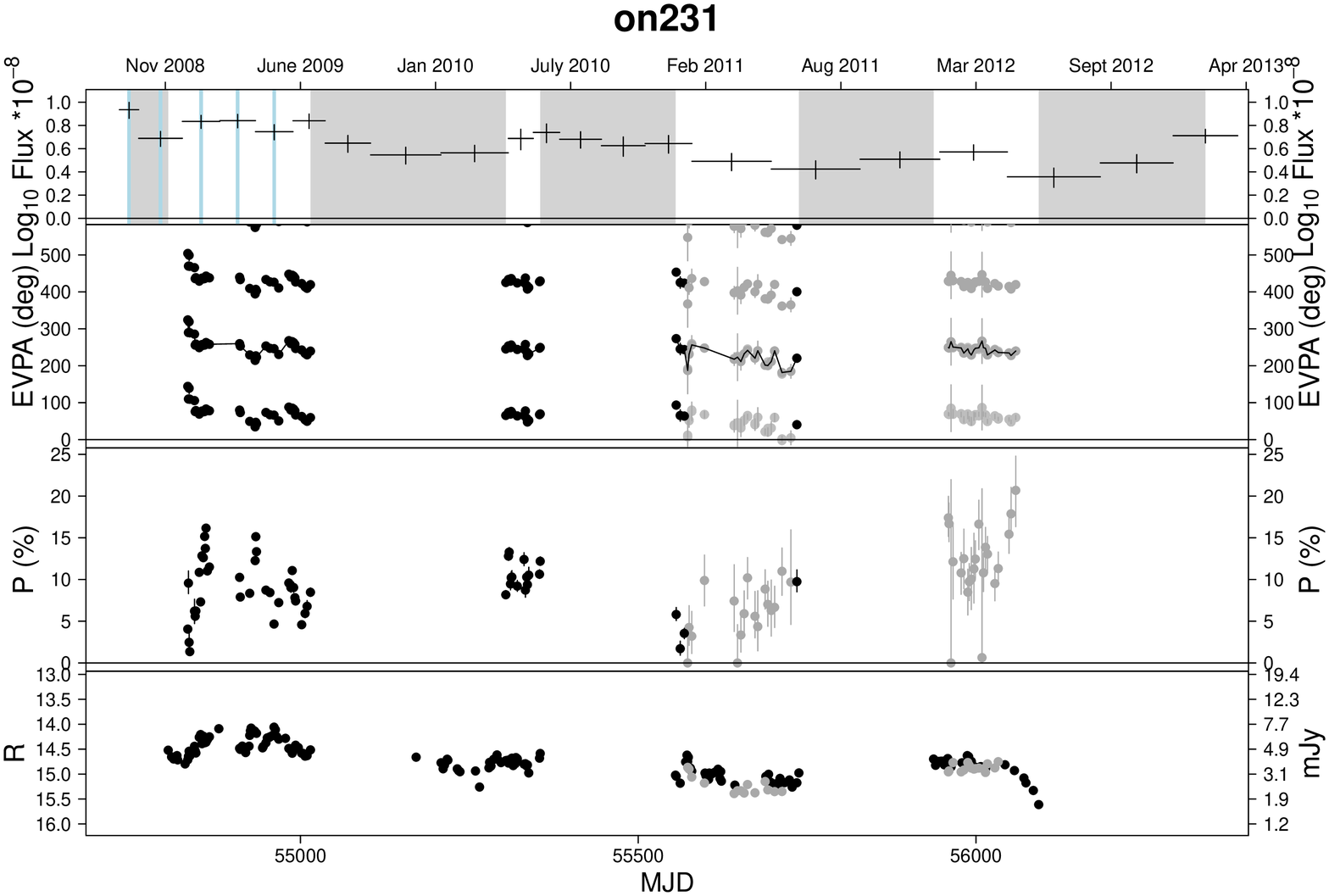}
\caption[on231_lcurves]{All $\gamma$-ray and optical data for ON~231. Top panel shows the Fermi $\gamma$-ray light curve. The errors on the x axis represent the bins used for the Fermi data. Grey vertical sections show periods where no synchronous optical data available. There are no flaring episodes identified in ON~231 during this period of time. The second panel shows the optical polarisation angle or electric vector position angle (EVPA), the grey points are RINGO2 data and the black points KVA-DIPOL data. The black line traces the temporally closest EVPA points, showing the most likely behaviour of the EVPA. The third panel shows the optical degree of polarisation, and the fourth panel the optical magnitude, all point colours are the same as those for panel 2.}
\label{fig:on231}
\end{figure}
\end{landscape}

\begin{landscape}
\begin{figure}
\centering
\includegraphics[width=22cm]{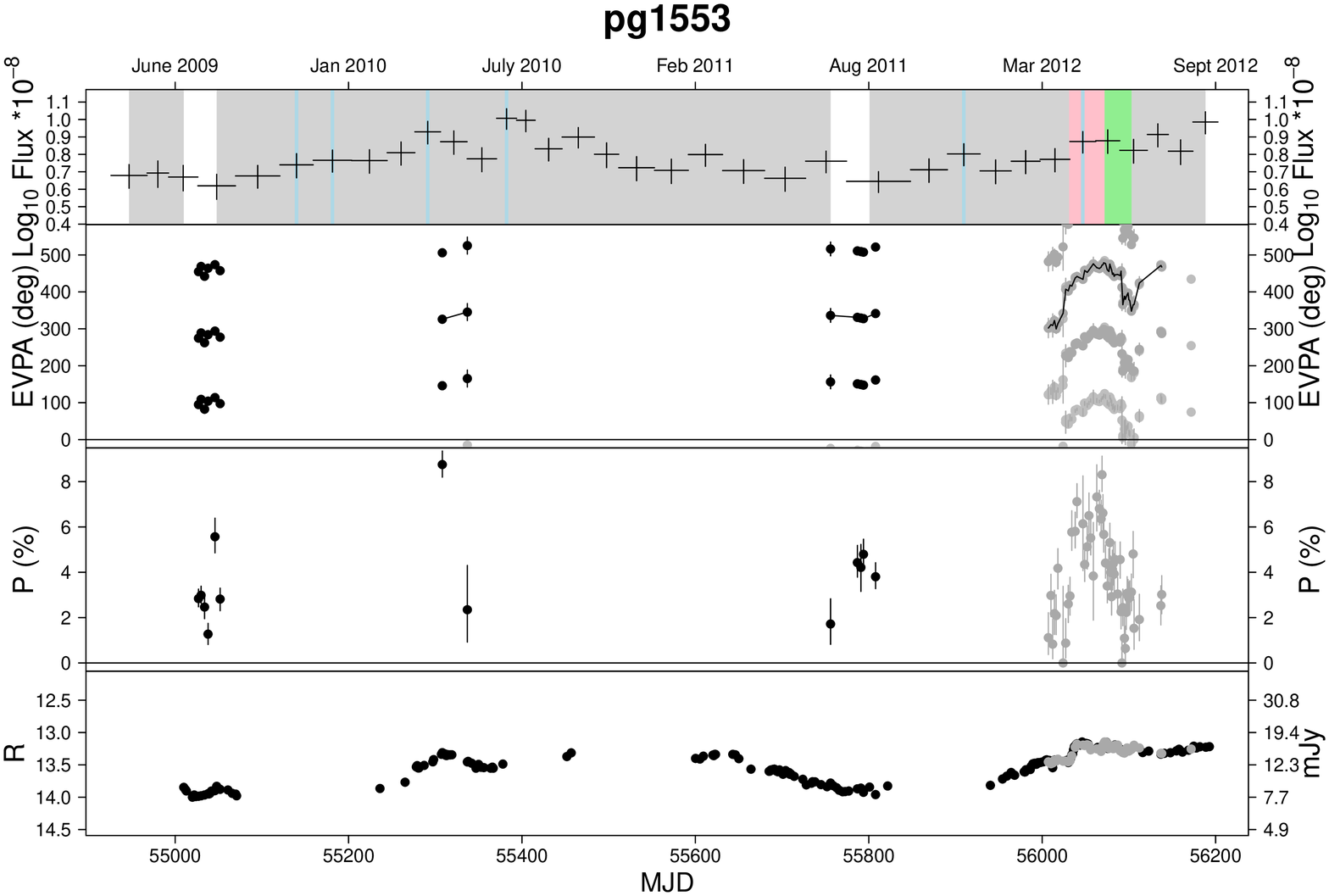}
\caption[pg1553_lcurves]{All $\gamma$-ray and optical data for PG~1553+113. Top panel shows the Fermi $\gamma$-ray light curve. The errors on the x axis represent the bins used for the Fermi data. Grey vertical sections show periods where no synchronous optical data available, pink vertical sections highlight regions where optical polarisation angle rotations occur in the upwards direction, light green sections show downward rotations. Flaring episodes are identified by vertical blue lines (see Section \ref{GRF} for definition of a flare). The second panel shows the optical polarisation angle or electric vector position angle (EVPA), the grey points are RINGO2 data and the black points KVA-DIPOL data. The black line traces the temporally closest EVPA points, showing the most likely behaviour of the EVPA. The third panel shows the optical degree of polarisation, and the fourth panel the optical magnitude, all point colours are the same as those for panel 2.}
\label{fig:pg1553}
\end{figure}
\end{landscape}

\begin{landscape}
\begin{figure}
\centering
\includegraphics[width=22cm]{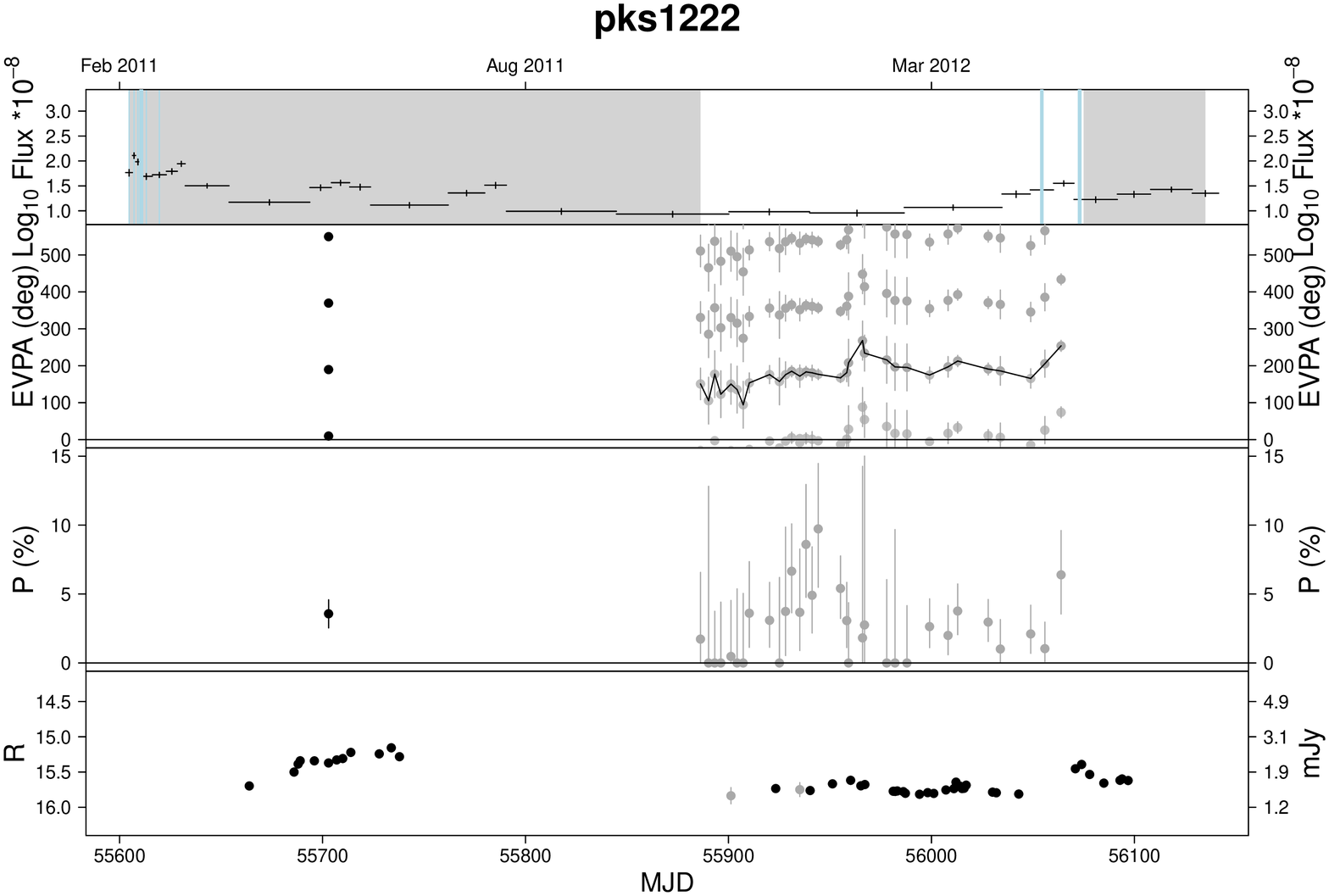}
\caption[pks1222_lcurves]{All $\gamma$-ray and optical data for PKS~1222+216. Top panel shows the Fermi $\gamma$-ray light curve. The errors on the x axis represent the bins used for the Fermi data. Grey vertical sections show periods where no synchronous optical data available. Flaring episodes are identified by vertical blue lines (see Section \ref{GRF} for definition of a flare). The second panel shows the optical polarisation angle or electric vector position angle (EVPA), the grey points are RINGO2 data and the black point is KVA-DIPOL data. The black line traces the temporally closest EVPA points, showing the most likely behaviour of the EVPA. The third panel shows the optical degree of polarisation, and the fourth panel the optical magnitude, all point colours are the same as those for panel 2.}
\label{fig:pks1222}
\end{figure}
\end{landscape}

\begin{landscape}
\begin{figure}
\centering
\includegraphics[width=22cm]{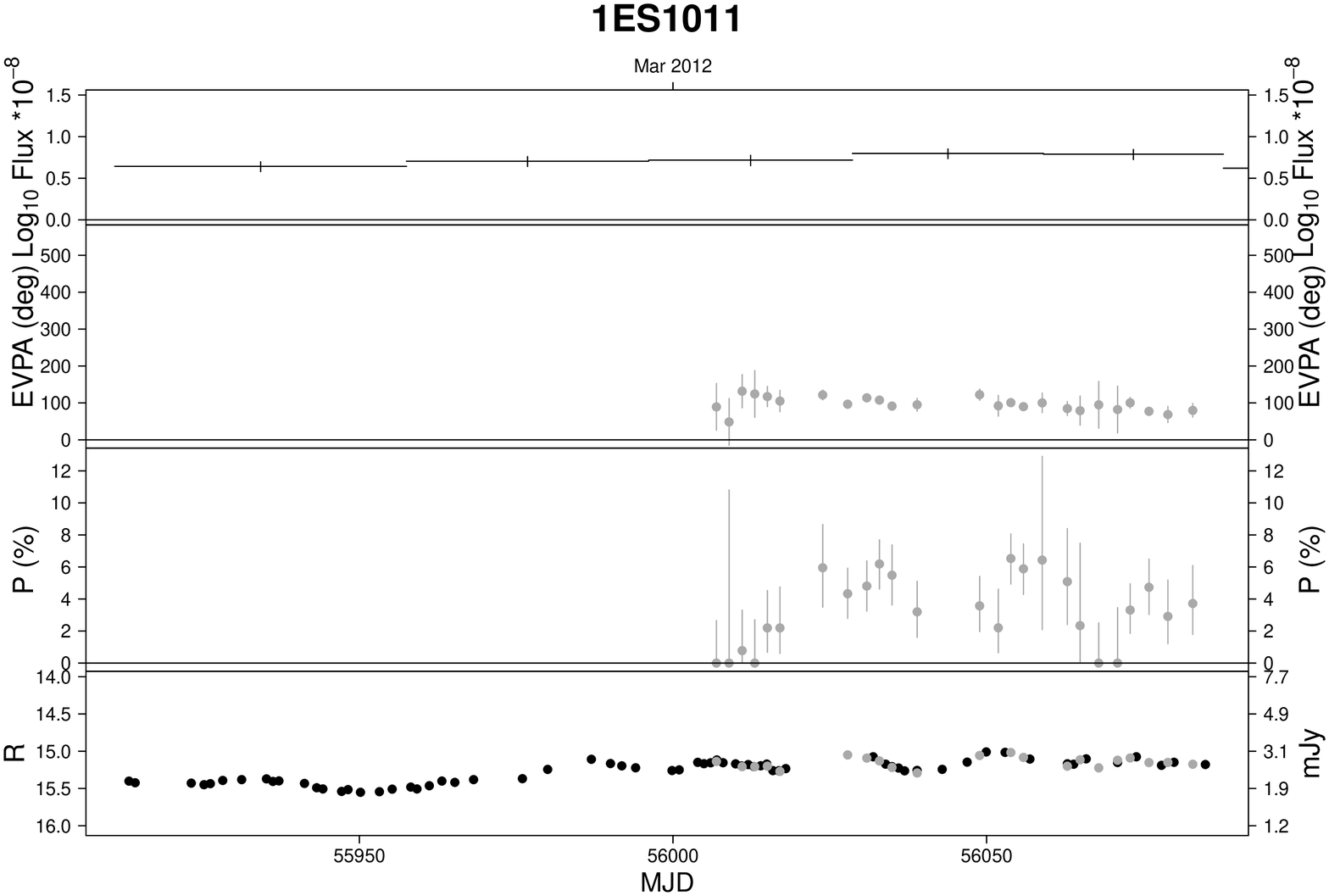}
\caption[1es1011_lcurves]{All $\gamma$-ray and optical data for 1ES~1011+496. Top panel shows the Fermi $\gamma$-ray light curve. The errors on the x axis represent the bins used for the Fermi data which, due to the faintness of the source, are quite large. No flare analysis was performed on this source due to the lack of Fermi data. The second panel shows the optical polarisation angle or electric vector position angle (EVPA), the grey points are RINGO2 data and no KVA-DIPOL data are available. The black line traces the temporally closest EVPA points, showing the most likely behaviour of the EVPA. The third panel shows the optical degree of polarisation, and the fourth panel the optical magnitude, all point colours are the same as those for panel 2.}
\label{fig:1es1011}
\end{figure}
\end{landscape}

\begin{landscape}
\begin{figure}
\centering
\includegraphics[width=22cm]{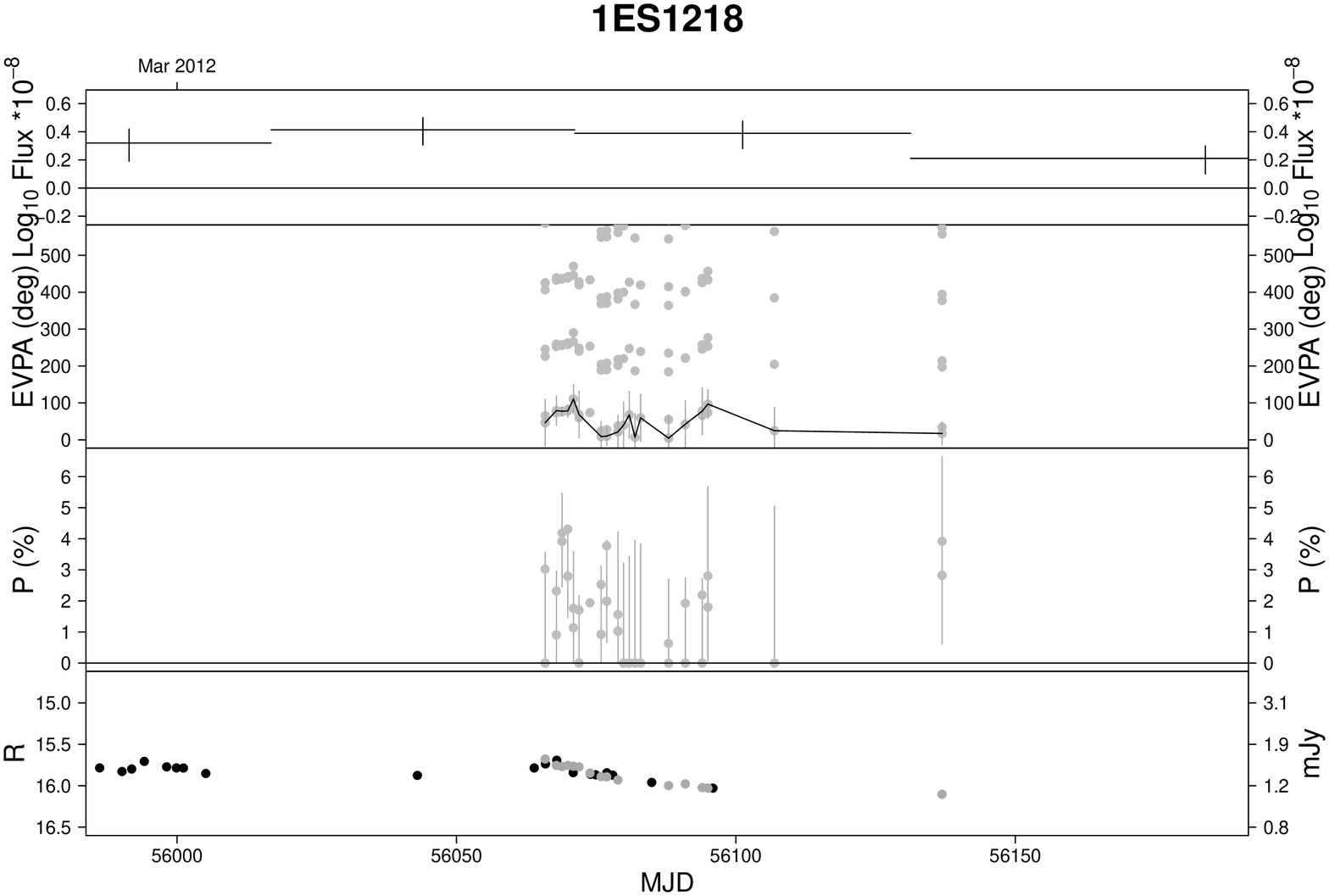}
\caption[1es1218_lcurves]{All $\gamma$-ray and optical data for 1ES~1218+304. Top panel shows the Fermi $\gamma$-ray light curve. The errors on the x axis represent the bins used for the Fermi data which, due to the faintness of the source, are quite large. No flare analysis was performed on this source due to the lack of Fermi data. The second panel shows the optical polarisation angle or electric vector position angle (EVPA), the grey points are RINGO2 data and no KVA-DIPOL data are available. The black line traces the temporally closest EVPA points, showing the most likely behaviour of the EVPA. The third panel shows the optical degree of polarisation, and the fourth panel the optical magnitude, all point colours are the same as those for panel 2.}
\label{fig:1es1218}
\end{figure}
\end{landscape}

\begin{landscape}
\begin{figure}
\centering
\includegraphics[width=22cm]{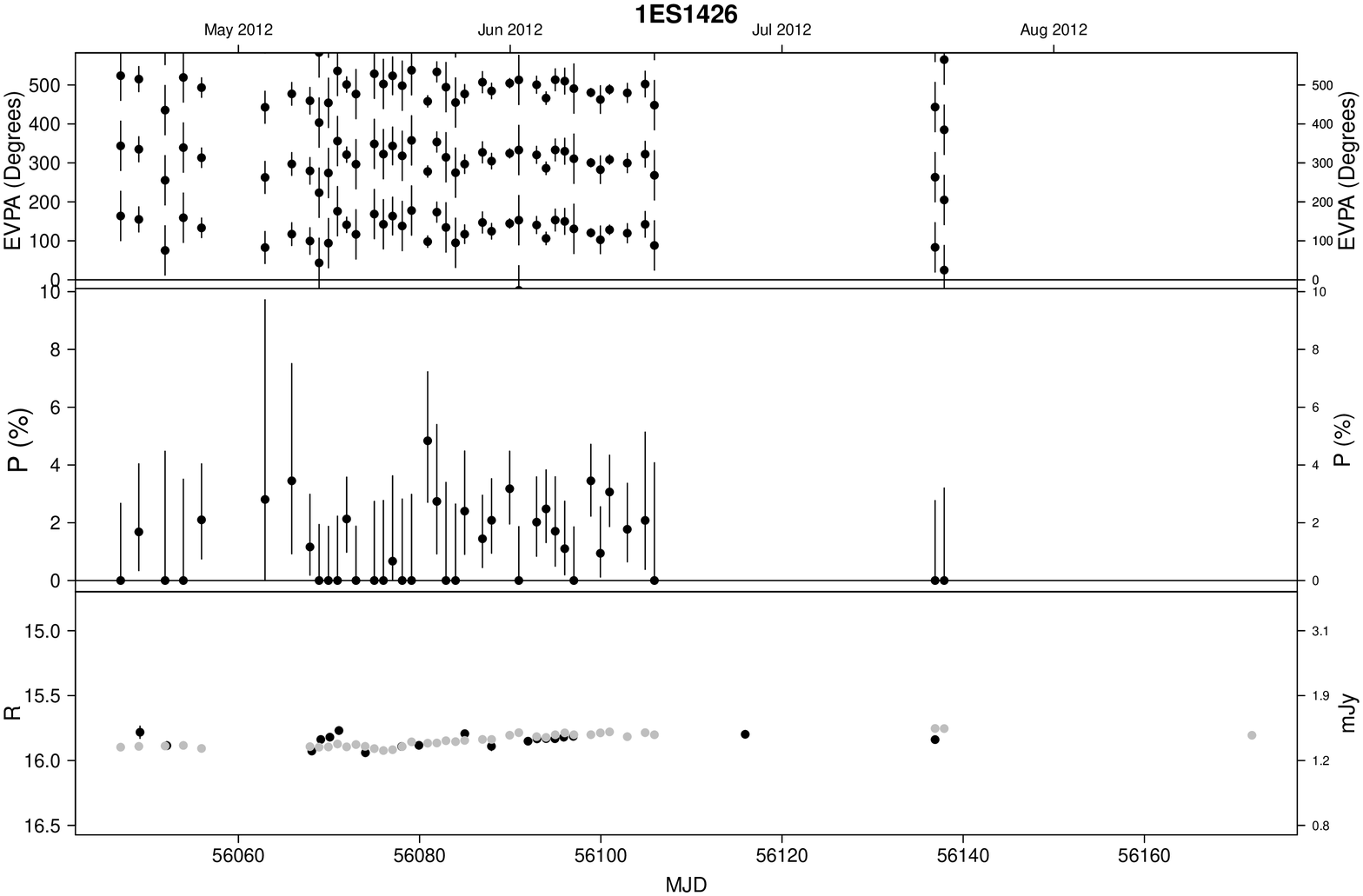}
\caption[1es1426_lcurves]{All optical data for 1ES~1426+428 (the source is too faint in Fermi data). Top panel shows the optical polarisation angle or electric vector position angle (EVPA), the black points are KVA-DIPOL data and there are no polarisation data points from RINGO2. The second panel shows the optical degree of polarisation, and the third panel the optical magnitude, all point colours are the same as those for Figure \ref{fig:mrk421}.}
\label{fig:1es1426}
\end{figure}
\end{landscape}
\label{LastPage}

\begin{table*}
\resizebox{0.82\textwidth}{!}{
\begin{tabular}{llllllll} \hline
\centering
Source/Season & MJD range                            & $\rho$ (gam-mag) & $\rho$ (gam-deg) & $\rho$ (deg-mag) & p  (gam-mag) & p  (gam-deg) & p (deg-mag) \\\hline
3C66A         &                                      			&               	&                   &               &              	&              &             \\\hline
a             & \textless55300                       		& 0.508         	& -0.287         & 0.299     & 7.30x10$^{-6}$            & 0.0659        & 0.00941       \\
b             & \textgreater55300 \& \textless 55700 & 0.0670         & -0.400         & ...           & 0.524        			& 0.000668        & ...         \\
c             & \textgreater55700 \& \textless 56100 & 0.119         	&  0.114        & ...           & 0.287           			& 0.261        & ...         \\
d             & \textgreater56100                    		& 0.491         	& 0.188        & ...           & 0.154        			& 0.608        & ...         \\\hline
S50716+714    &                                      		&           	    	&                    &               &              	&                  &             \\\hline
a             & \textless55000                       		& 0.480         	& -0.0129        & -0.0754   & 9.94x10$^{-6}$           & 0.946        & 0.759        \\
b             & \textgreater55000 \& \textless55400  & 0.523         	& 0.411        & ...           & 2.54x10$^{-5}$          & 0.101        & ...         \\
c             & \textgreater55400 \& \textless 55800 & 0.127         	& 0.272        & ...           & 0.302        			& 0.0108        & ...         \\	
d             & \textgreater55800                    		& 0.586         	& -0.0324        & ...           & 1.83x10$^{-6}$             & 0.758        & ...         \\\hline
OJ287         &                                      			&               	&                   &               &              	&                  &             \\\hline
a             & \textless55000                       		& 0.493          & -0.230          & 0.000547     & 3.41x10$^{-6}$            & 0.177        & 0.999        \\
b             & \textgreater55000 \& \textless55400  & 0.565         & -0.235        & ...           & 3.75x10$^{-6}$            	& 0.331        & ...         \\
c             & \textgreater55400 \& \textless55800  & -0.304         & 0.351        & ...           & 0.0153       			& 0.120        & ...         \\
d             & \textgreater55800                    		& 0.470         & 0.0126        & ...           & 1.54x10$^{-4}$          	& 0.925         & ...         \\\hline
1ES1011+496 	&                                      		&                   &               	    &               &              		&                  &             \\\hline
a             & \textgreater56000                    		& ...               & ...           	    & 0.500       & ...          		& ...              & 0.0191        \\\hline
Mrk421        &                                      			&                   &               	    &               &              		&                  &             \\\hline
a             & \textgreater55800 \& \textless55150  & 0.577   		& 0.0649		& 0.0933     	& 2.2x10$^{-16}$        & 0.296         & 0.133   \\\hline
Mrk180        &                                      			&                	&               	&              	&              	&                  &             \\\hline
a             & \textless56025                       		& ...           	& ...           	& 0.429          	& ...          	& ...          & 0.419        \\\hline
1ES1218+304   &                                      		&               	&               	&               	&              	&              &             \\\hline
a             & \textgreater56050                    		& ...           	& ...           	& 0.111          & ...          		& ...          & 0.695         \\\hline
ON231         &                                     		        &               	&               	&               	&              	&              &             \\\hline
a             & \textless55100                       		& 0.0860         & 0.130         & 0.395     & 0.506        			& 0.455    & 1.53x10$^{-11}$        \\
b             & \textgreater55100 \& \textless55500  & 0.297          & 0.126         	& ...           & 0.0660        		& 0.683     & ...         \\
c             & \textgreater55500 \& \textless55800  & 0.635          & -0.560         	& ...           & 1.64x10$^{-4}$         & 0.00830            & ...         \\
d             & \textgreater55800                    		& 0.708          & -0.305         	& ...           & 7.43x10$^{-5}$        	& 0.178     & ...         \\\hline
PKS1222+216   							&                   &               	&               	&               	&              &              &             \\\hline
a             & \textless55800                       		& 0.0303         & ...           	& 0.245    & 0.946        	& ...          & 0.342        \\
b             & \textgreater55800                    		& -0.600         & -0.199         	& ...           & 8.23x10$^{-6}$        	& 0.174    & ...         \\\hline
3C279         &                                      			&               	&               &               	&              	&              &             \\\hline
a             & \textless55100                       		& 0.711           & -0.249     & 0.270     	& 0.000            & 0.263   & 0.0607       \\
b             & \textgreater55100 \& \textless 55500 & 0.376         & ...           & ...           	& 0.0238        	& ...          & ...         \\
c             & \textgreater55500 \& 55800           	& 0.697         & -0.0197        & ...           & 6.54x10$^{-9}$            & 0.912     & ...         \\
d             & \textgreater55800                    		& 0.489         & 0.556        & ...           & 0.00525            & 1.52x10$^{-6}$     & ...         \\\hline
1ES1426       &                                      			&               	&               &               	&              	&              &             \\ \hline
a             & \textless56120                       		& ...           	& ...           & 0.105         & ...          	& ...          & 0.518       \\\hline
PKS1510-089   &                                      		&               	&               &             	&              	&              &             \\\hline
a             & \textless55100                       		& 0.669          & 0.360        & 0.0986      & 3.67x10$^{-12}$        		& 0.0248    & 0.486        \\
b             & \textgreater55100 \& \textless55500  & 0.00339         & ...           & ...           	& 0.988        				& ...          & ...         \\
c             & \textgreater55500 \& \textless55900  & 0.554         & -0.0277        & ...           	& 4.51x10$^{-7}$            & 0.817   & ...         \\
d             & \textgreater55900                    		& 0.565         & 0.362        & ...           	& 1.17x10$^{-8}$            	& 3.91x10$^{-5}$   & ...         \\\hline
PG1553+113    &                                      		&               	&               &               	&              	&              &             \\\hline
a             & \textless55100                       		& 0.180        & 0.419      & 0.549          & 0.435        				& 0.419    & 1.98x10$^{-5}$           \\
b             & \textgreater55100 \& \textless 55500 & 0.745         & ...           & ...           	& 2.34x10$^{-6}$        		&	 ...           & ...         \\
c             & \textgreater55500 \& \textless 55900 & 0.0718         & ...           & ...           	& 0.620          	&...            	& ...         \\
d             & \textgreater55900                    		& 0.502          & 0.633        & ...           & 5.60x10$^{-16}$            	& 1.89x10$^{-13}$     & ...         \\\hline
Mrk501        &                                      			&               	&               &               	&              &              &             \\\hline
a             & \textless55900                       		& 0.109           & -0.121         	& 0.0876      	& 0.262        			& 0.235     & 0.403        \\
b             & \textgreater55900                    		& -0.0929         & 0.160        & ...           	& 0.152         			& 0.0288      & ...         \\\hline
BL Lac        &                                      			&               	&               &              	 &              &              &             \\\hline
a             & \textless55300                       		& 0.287         & -0.210         & -0.485    & 0.000493        			& 0.0627        	& 1.50x10$^{-6}$           \\
b             & \textgreater55300 \& \textless55650  & 0.680         & 0.0168        & ...           & 4.97x10$^{-11}$            	& 0.900        	& ...         \\
c             & \textgreater55650 \& \textless55990  & 0.548         & -0.205         & ...           & 1.71x10$^{-5}$            	& 0.0719            & ...         \\
d             & \textgreater55990                    		& 0.718         & 0.143          & ...           & 3.66x10$^{-8}$        	& 0.803          	& ...        \\ \hline

\end{tabular}}
\caption[]{Full table of Spearman Rank Correlation results ($\rho$ and p value) for all sources and all seasons. Empty fields represent sources that were not observed in that season or lack one of the datasets required for that analysis and for the optical magnitude and degree of polarisation values there is only one for each source as the data were not split into seasons (see Section \ref{optopt} for more details).}
\label{full_table}
\end{table*}

\end{document}